\documentclass[twocolumn,showkeys,preprintnumbers,amsmath,amssymb]{revtex4} 

\usepackage{graphicx}% Include figure files
\usepackage{dcolumn}% Align table columns on decimal point
\usepackage{bm}% bold math
\usepackage{bbm}
\usepackage{amsmath}
\usepackage{mathtools}

\begin{document}

\title{Self-sustained optomechanical state destruction triggered by the Kerr nonlinearity}

\author{A.$~$Delattre$^{*}$, I.$~$Golokolenov$^{*}$, R. Pedurand$^{*}$, X.$~$Zhou$^{**}$, A.$~$Fefferman$^{*}$ and E.$~$Collin$^{*,\dag}$}

\address{(*) Univ. Grenoble Alpes, Institut N\'eel - CNRS UPR2940, 
25 rue des Martyrs, BP 166, 38042 Grenoble Cedex 9, France \\
          (**) IEMN, Univ. Lille - CNRS UMR8520, 
Av. Henri Poincar\'e, Villeneuve d'Ascq 59650, France 
 }

\date{\today}

\begin{abstract}
Cavity optomechanics implements a unique platform where moving objects can be probed by quantum fields, either laser light or microwave signals. 
With a pump tone driving at a frequency above the cavity resonance, self-sustained oscillations can be triggered at large injected powers. 
These limit cycle dynamics are particularly rich, presenting hysteretic behaviours, broad comb signals and especially large motion amplitudes. 
All of these features can be exploited for both fundamental quantum research and engineering.
Here we present low temperature microwave experiments performed on a high-Q cavity resonance capacitively coupled to the flexure of a beam resonator.  
We study the limit cycle dynamics parameter space as a function of pump parameters (detuning, power).
Unexpectedly, we find that in a region of this parameter space the microwave resonance is irremediably destroyed: only a dramatic power-reset can restore the dynamics to its original state.
The phenomenon can be understood as an optical instability linked to the Kerr nonlinearity of the cavity. A theory supporting this claim is presented, reproducing almost quantitatively the measurement. 
This remarkable feature might be further optimized and represents a new resource for quantum microwave circuits.

\end{abstract}


\maketitle
%\vspace*{-0.2mm}

%
\section{Introduction}

Nanomechanical devices embedded in microwave cavities are the expression of cavity optomechanics obtained when the light field frequency is decreased down to the GHz domain. The Fabry-P\'erot cavity is then replaced by an $RLC$ resonator, and the light-matter interaction is created by capacitive coupling: the motion $x$ of an electrode modulates the capacitance $C$ of the microwave resonator, changing thus its frequency $\omega_c$ \cite{regal2008}.
These systems inherit the properties of optomechanical platforms, are easier to cool down and can be readily interfaced with quantum electronics circuits.
Cooling down the mechanical mode to near its ground state of motion offers then unique possibilities where scientists can study the foundations of quantum physics: e.g. measuring the imprint of zero point fluctuations \cite{schwabPRX2014} or entangling the flexural modes of distinct mechanical objects \cite{sillanpaaEntangle}.

Beyond fundamental research, microwave optomechanics can be thought of as a resource for quantum electronics and quantum information processing. For instance, quantum-limited non-reciprocal microwave elements \cite{nonrec} and visible-to-microwave photon converters \cite{photonsConv} are being developed.

A rather specific dynamical regime can be reached when a microwave pump tone is applied to the system at a frequency $\omega_p$ about $\omega_c+\Omega_m$, with $\Omega_m$ the mechanical resonance frequency. This so-called Stokes pumping scheme generates anti-damping on the mechanics, and with strong enough drive powers $P_{in}$ one can trigger self-sustained oscillations \cite{AKMreview}.
This regime is characterized by multiple attractor states that generate bifurcations and hystereses \cite{marquardtBif}. A specificity of these states is to imply very large motion and optical field amplitudes, with extremely narrow spectral lines organized in a comb \cite{DylanPRR}.
Such states, beyond the profound understanding that they deserve, can be used as unique tools e.g. for sensing applications. Indeed, a recent work applied self-oscillation optomechanics to force sensing \cite{FaveroForce}; which might be extended in the future to the more specific case of mass spectrometry \cite{natcommMasssensing}. Besides, these large motion amplitude states can be utilized to quantitatively characterise the nonlinearities in the optomechanical coupling \cite{DylanPRR}.
This simple, in-built capability could be specifically used in dedicated experiments focused on quantum gravity, for which the knowledge of these nonlinearities is essential \cite{quantgravitnonlin}.

In the present paper, we report on experiments performed on a beam-based microwave optomechanical setup at cryogenic temperatures. We construct the $(P_{in}, \delta)$ parameter space of the self-oscillation regime, with $P_{in}$ the microwave pump power injected on chip, and $\delta$ the pump frequency shift from the ideal detuning of $\omega_c+\Omega_m$. What we primarily measure is the output photon field, which consists in a comb of peaks from which we extract amplitudes  and frequencies. These are reproduced here in terms of maps of the photon amplitude of the main peak (located around $\sim \omega_c$), and of mechanical frequency shifts (with respect to the Brownian value $\Omega_m$), in a similar manner to Ref. \cite{DylanPRR}. 

%%%% now the point:
The observed parameter space is rather different from what would be expected from linear theory, and more importantly {\it does not match} the features presented in Ref. \cite{DylanPRR}.
The measured amplitude is an order of magnitude weaker than expected, and presents a specific 
%%%%  MODIF!
 non-symmetric shape with respect to frequency detuning $\delta$.
%%% {\it moir\'e} contrast. 
%%%%
Interestingly, the hysteretic feature observed in Ref. \cite{DylanPRR} appears {\it at a different place} of the parameter space.
And finally, the most remarkable finding of these measurements is a region where the combined opto-mechanical dynamics is simply {\it destroyed}: the comb peaks disappear, the cavity susceptibility cannot be seen anymore, no matter how we detune the pump tone $\delta$ or change its power $P_{in}$ within the parameter space. {\it Only} by decreasing substantially $P_{in}$ and coming back to the stable region can we recover the optomechanical dynamics, in a reproducible manner. This is what we shall call a "dramatic" power-reset.

%%%% And intro
To understand these intriguing features, one has to introduce some nonlinear ingredients. Since we do not match Ref. \cite{DylanPRR}, it is obvious that nonlinear coupling will not be enough. The mechanical Duffing effect which arises from stretching under motion \cite{duffingbook} cannot influence the optical field that much either \cite{DylanPRR}. 
We therefore invoke the Kerr nonlinearity of the superconducting cavity, which was safely neglected in Ref. \cite{DylanPRR} because the cavity had a poor Q factor and the good coupling strength $g_0$ did not require that large microwave fields to be used.
On the other hand, the situation is reversed for the device studied here: we have a very large Q and a weak $g_0$. 

%%%%%%%%%%
We present a theoretical model that reproduces the  measured characteristics, and we discuss its implications. Our main claim is that the Kerr terms in the Hamiltonian are responsible for a {\it bifurcation in the optical field}, which must be toward a state of very large photon population.
We propose that this population is enough to destroy the superconductivity in the cavity, which extinguishes the optomechanical state. Only by decreasing drastically the injected power can we restore superconductivity, and in turn the optomechanical dynamics. 
 %%%% ADDD
The observed features depend strongly on the Kerr effect, which can be fit on data. Its knowledge 
is particularly relevant to optomechanics experiments, since it can modify basic signatures such as sideband asymmetry \cite{tobias_SB}.
%%% ADDED REFS!
Furthermore, the Kerr nonlinearity appears as a new resource for optomechanics, as already suggested in the framework of ground-state cooling schemes \cite{metelmann} or transduction amplification \cite{haviland, soliton}; an approach than can be found also in related topics as magnomechanics \cite{davis} and superconducting circuits \cite{wiebke}. 
%%%%%
The nonlinearity could then be tailored (e.g. by using granular Al as superconductor \cite{popAl}, or inserting a Josephson junction \cite{agustin}) in order to make use of the power-reset effect in quantum electronics.

\section{Experiment}
\label{expt}

\begin{figure}[t!]
		\centering
	\includegraphics[width=8.5cm]{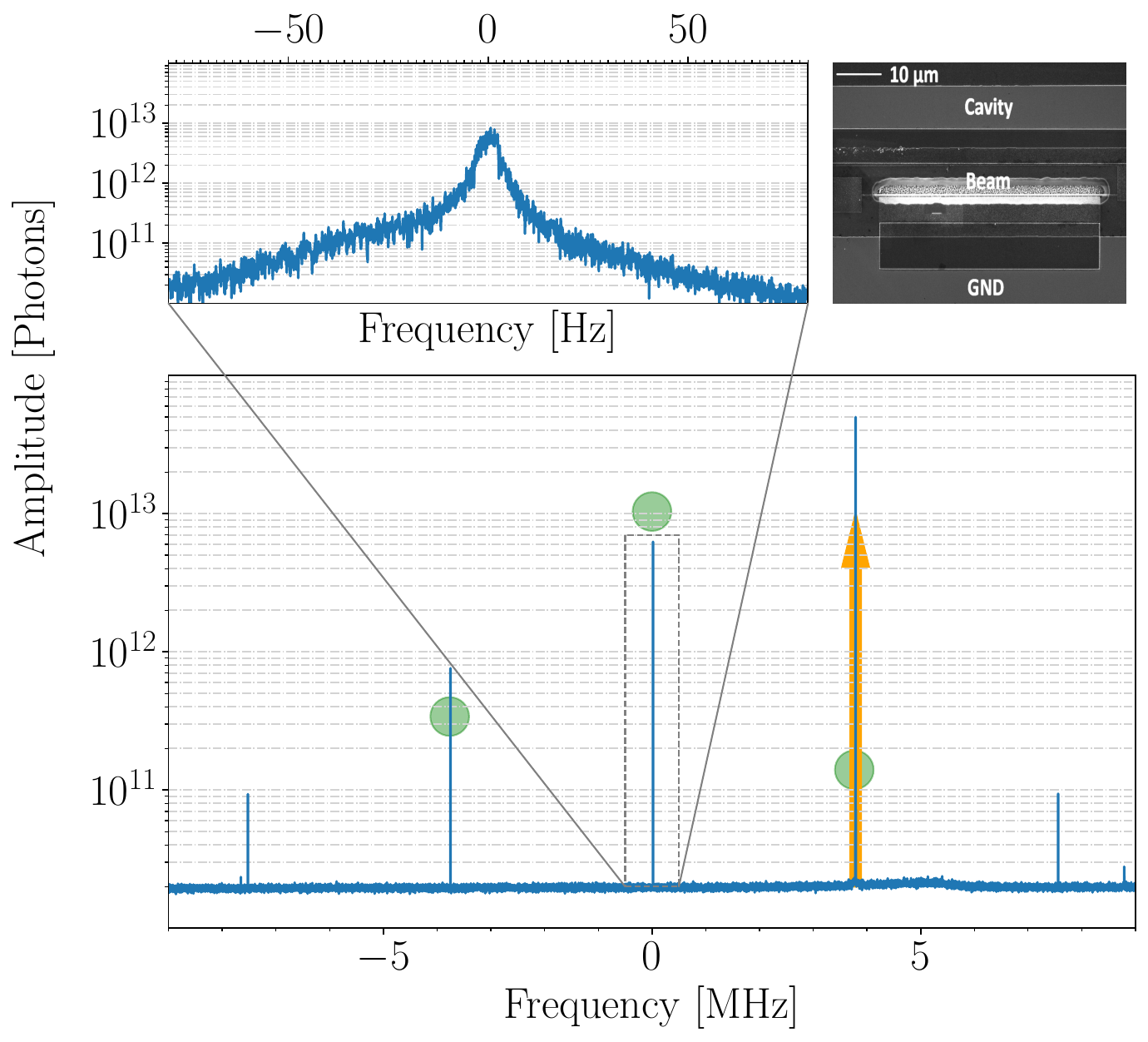}
	\vspace*{0.5cm}
			\caption{
			%(Color online) 
			Power Spectral Density (PSD, in photons of energy $\hbar \omega_c$) measured in the self-sustained oscillation regime at 400$~$mK (pump power $P_{in}=130~$nW and detuning $\delta/(2\pi)=0~$Hz). The acquisition BW here was 0.1$~$Hz, much smaller than the one used to scan the parameter space. Left inset: zoom-in on the peak at frequency $\sim \omega_c$. Right inset: scanning electron microscope (SEM) image of the beam device. Green dots are from the Kerr theory, and the arrow indicates the position of the pump tone, where the large injected power adds to the output field (see text). }
			\label{fig1}
\end{figure}

The device studied here is the same as the one originally used in Ref. \cite{XinPRAppl,XinErratum}. 
It consists in a coplanar $50~\Omega$ $\lambda/4$ Nb cavity resonating at $\omega_c = 2 \pi \times 5.988~$GHz, coupled to the $\Omega_m =2 \pi \times 3.8~$MHz first flexure of an Al-covered SiN beam mechanical element (nominally $50~\mu$m $\times 300~$nm $\times 100~$nm, see Fig. \ref{fig1} right inset). The single photon/phonon coupling strength is about $g_0 \approx 2 \pi \times 0.5~$Hz.
 Microwaves are fed in/out the cavity with a coupling rate of approximately $\kappa_{ext} = 2\pi \times 80~$kHz, for a total decay rate of about $\kappa_{tot} = 2 \pi \times 150~$kHz. This makes the microwave resonator reasonably high-Q. The experiment is performed in reflection mode, as in Ref. \cite{IlyaPRRthermo}, although the design of the coupling is bidirectional. 
%%%
Microwave amplitudes have been calibrated to within $\pm 2~$dB (absolute) on both injection and detection, see Ref. \cite{XinPRAppl}.
%%% 
 Measurements are performed in a commercial dilution cryostat at hundreds of milliKelvin temperatures. Apart from small frequency shifts in the resonances and thermal noise, the only parameter that depends significantly on temperature $T$ is the mechanical relaxation rate $\Gamma_m$, essentially $ \propto T$.
At 400$~$mK, it is about 40$~$Hz. Potential heating effects due to high microwave powers are discussed in Appendix \ref{HeatingandT}.
 
\begin{figure}[t!]
		\centering
	\includegraphics[width=9.cm]{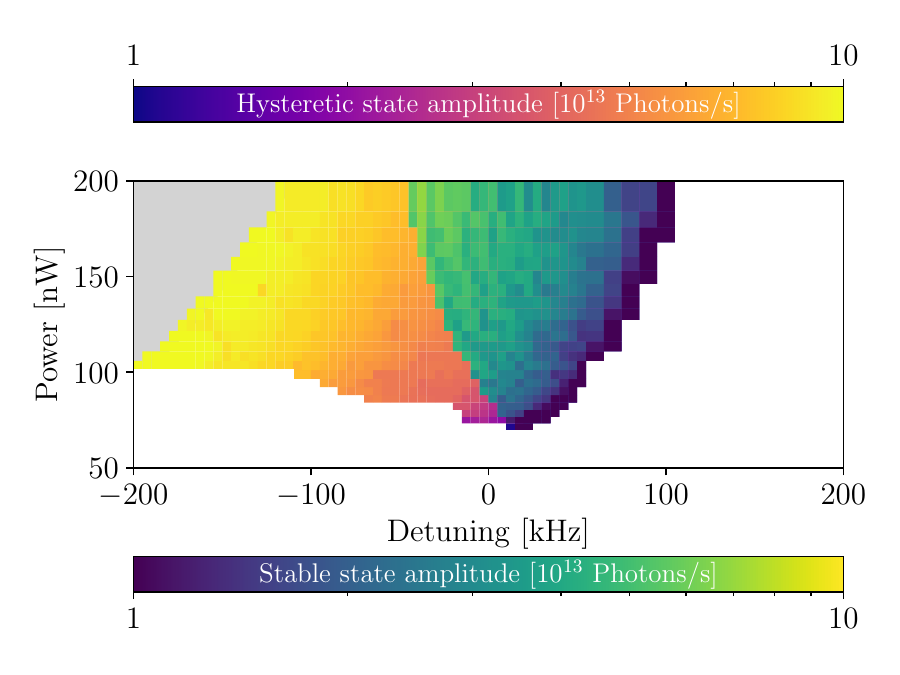}
	%\vspace*{0.5cm}
			\caption{
			%(Color online) 
			Measured self-oscillating amplitude of the main comb peak (at $\sim \omega_c$) as a function of drive parameters $(P_{in},\delta)$ at 400$~$mK. The green colour stands for the stable zone, the yellow  for the hysteretic one, while the grey background represents the region where the signal is lost, and cannot be recovered without a power-reset.
			%%%% disclaimer
Note that in the hysteretic region where we do observe the coexistence of two states (self-oscillating or Brownian motion), the attractor in which the system resides is nevertheless stable (see text for details). }
			\label{fig2}
\end{figure}

We apply a single pump tone at frequency $\omega_p=\omega_c+\Omega_m+\delta$, of power $P_{in}$, and measure the output microwave spectrum for given $(P_{in},\delta)$. 
The measurement is performed with the same setup as in Ref. \cite{DylanPRR}, in which a high-frequency lock-in amplifier is used to extract the Power Spectral Density (PSD). 
When self-oscillation starts, it consists of a comb of very narrow non-Lorentzian peaks, about a few Hz wide at $\sim 6~$GHz, as shown in Fig. \ref{fig1}. 
Their width simply reflects the phase noise experienced by the resonances (see left inset).
Note that fewer peaks are visible compared to Ref. \cite{DylanPRR}, simply because the sideband-resolved ratio $\kappa_{tot}/\Omega_m$ is smaller here. 
Scanning the parameter space in $(P_{in},\delta)$, we show in Fig. \ref{fig2} the resulting amplitude of the main comb peak, the one located at $\sim \omega_c$ (the $x$-axis $0$ in Fig. \ref{fig1}).
%%%
%%%% XXXX
The acquisition is then performed with a bandwidth (BW) much larger than the peak width, leading to the integration of the peak signal (in photons/second).
%%%
From its position, we extract the mechanical resonance frequency analysed in Appendix \ref{DuffingShiftg1g2}.
 The color-coding highlights the different regions discussed in the introduction: stable zone in green, hysteretic one in yellow and 
"destroyed" comb in grey.

%%%% ADDED!
Because of slight irreproducibilities in the measured %wiggly 
pattern shown in Fig. \ref{fig2} (see Appendices \ref{HeatingandT} and \ref{DuffingShiftg1g2}), the measurement protocol has been kept as simple as possible, such that scanning the parameter space is reasonably fast (it takes a day), and can be done with a minimum amount of stitchings: the power is slowly ramped up from the Brownian regime (very low powers) around detuning $\delta/(2\pi) \sim 0~$Hz to a fixed value $P_{in}$ (within the colored region of Fig. \ref{fig2}), and then $\delta$ is either swept down or up until self-oscillation stops. This is then repeated from the lowest input power to the highest, creating two half spaces which are concatenated. If everything goes on well, the resulting pattern is reasonably continuous over the full parameter space (Fig. \ref{fig2}). %However when the full map is created from many portions, the overall picture can be a bit messy 
Irregularities in the measured pattern are discussed in 
 Appendix \ref{HeatingandT}, together with the temperature dependence (Fig. \ref{figA}).
%%%% 

 When we scan the parameter space ramping the power up at different fixed detunings $\delta$, starting from the lowest negative values and increasing it to the positive ones, the yellow region in Fig. \ref{fig2} is never triggered: this is the signature of hysteresis (Appendix \ref{MechHysteresisSolve}). 
%%%% ADD Andrew
Besides, the grey region is triggered only when arriving from the yellow zone (that is, from the self-oscillation state). Then, self-oscillation stops and no microwave signal comes out of the cavity anymore: the cavity susceptibility cannot be found from a probe frequency sweep. Bringing then $(P_{in},\delta)$ back into the green region, no self-oscillation can be seen either, and a power-reset is required to re-activate the system. This is in drastic contrast with what happens at positive detunings: there, when self-oscillation stops the cavity susceptibility remains visible, and self-oscillation can be triggered again by sweeping $\delta$ back into the green zone.
%Describe here the different behavior at negative and positive detuning when the self-oscillation stops. In particular, specify that for negative detuning, when you reach the grey region, (1) the cavity doesn't appear in a probe spectrum and (2) self-oscillation is not recovered when the detuning is increased toward zero at constant power. These two features are absent when the self-oscillation stops at positive detuning.
%%%%%% 
Finally, note the straight "staircase-like" bottom of this yellow zone which is due to our sweeping technique \cite{DylanPRR}, discussed in Appendix \ref{DuffingShiftg1g2}. % and Section \ref{results} below.
%%%%

\begin{figure}[t!]
		\centering
	\includegraphics[width=9.cm]{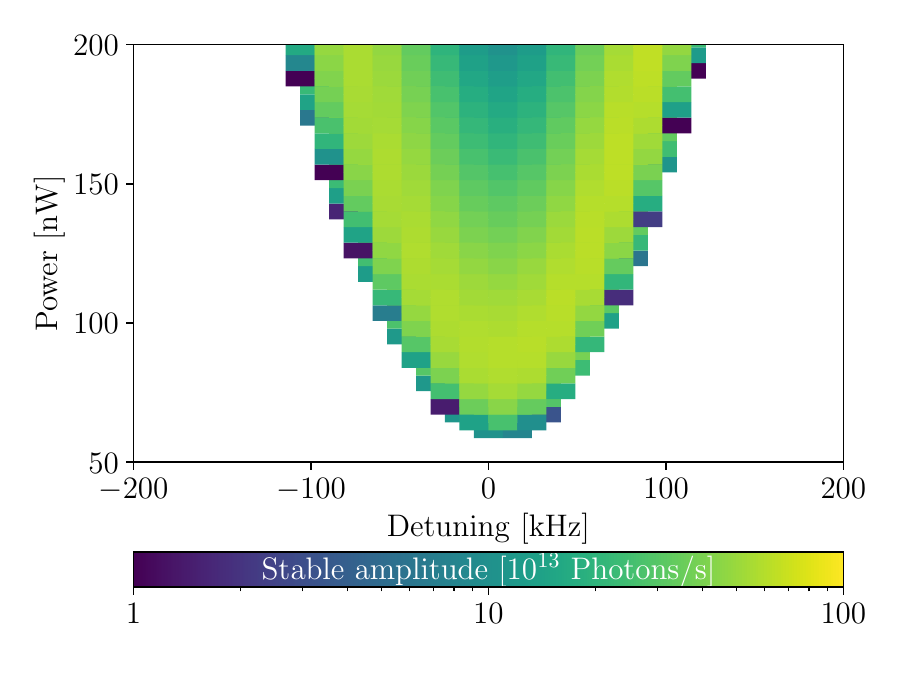}
	%\vspace*{-0.5cm}
			\caption{
			%(Color online) 
			Calculated plot corresponding to Fig. \ref{fig2} from the linear theory \cite{CRphysiqueADA}. Note that the scales of the two graphs {\it differ by an order of magnitude} (see text). }
			\label{fig3}
\end{figure}

For comparison, we present in Fig. \ref{fig3} the expected self-oscillation map calculated from the linear theory.
Figs. \ref{fig2} and \ref{fig3} look dramatically different. Besides, it is impossible to generate an instability such as the grey zone in these plots with only mechanical nonlinearities, such as the ones analyzed in Ref. \cite{DylanPRR}. We therefore {\it must} involve optical field related nonlinear coefficients. These Kerr type terms are analyzed in Section \ref{theory} below. 
However, nonlinear coupling and Duffing effects are obviously also present in this device. Their additional imprint is discussed in Appendix \ref{DuffingShiftg1g2} and Section \ref{results}. %; the nonlinear coupling terms proper fit is clearly outside of the scope of this manuscript. %, they are nonetheless commented in Section \ref{results}.	

\section{Theory}	
\label{theory}

We investigate the influence of Kerr nonlinear terms on the dynamics of self-sustained oscillations. 
The generic Hamiltonian which is considered here is given in Appendix \ref{HamiltonRWT}.
Thanks to the wide separation of time-scales and  weak coupling and damping, we can treat the problem self-consistently introducing a slowly changing mechanical amplitude  \cite{marquardtBif,CRphysiqueADA,Dykman1}.
In the rotating frame of the pump optical field, the Hamiltonian writes:
\begin{eqnarray}  \label{hamilton}
%\begin{gathered}  
\hat{H} & = &-\hbar\left[ \Delta'+g_{0}\left(\hat{b}+{\hat{b}}^{\dag}\right) 
\right]{\hat{a}}^{\dag}\hat{a}\\ \nonumber 
& + & \hbar \Lambda_1 \left( {\hat{a}}^{\dag}\hat{a} \right) \left( {\hat{a}}^{\dag}\hat{a} \right)+\hbar \Lambda_2 \left( {\hat{a}}^{\dag}\hat{a} \right) \left( {\hat{a}}^{\dag}\hat{a} \right) \left( {\hat{a}}^{\dag}\hat{a} \right)  \\ \nonumber
& +& \hbar\Omega_{m} {\hat{b}}^{\dag}\hat{b}
-\mathbbm{i} \hbar \sqrt{\frac{\kappa_{ext}}{2}} \left[ {\hat{a}}^{\dag} \tilde{\alpha}_{p} e^{-\mathbbm{i}\varphi_p} -\hat{a}\, \tilde{\alpha_p}^{*}e^{+\mathbbm{i}\varphi_p} \right] , 
%\end{gathered}
\end{eqnarray}
keeping the two lowest order Kerr terms $\Lambda_1,\Lambda_2$.
Here $\hat{a}$ and $\hat{b}$ are the photon and phonon annihilation operators, respectively. 
Note the $\kappa_{ext}/2$ which comes from the bidirectional design of the chip \cite{XinPRAppl}; for a single-port device, remove the $1/2$ (see input-output theory) \cite{devoret}.
The drive field is defined as $\alpha_p (t) = \tilde{\alpha}_{p} \, \mathrm{exp} [-\mathbbm{i} (\omega_p t + \varphi_p)]$, with amplitude $\tilde{\alpha}_{p}$ and phase $\varphi_p$. We introduce the pump detuning parameter $\Delta = \omega_p-\omega_c=\Omega_m+\delta$, which for ideal Stokes pumping should verify $\delta =0$. 
%%%% introduce now!
By definition, $P_{in}= \hbar \omega_p \,  \tilde{\alpha}_{p}^2 $.
The prime in Eq. (\ref{hamilton})
comes from the definition of $\Delta'$ using a  cavity resonance $\omega_c'$ slightly renormalized by the Kerr terms, see Appendix \ref{HamiltonRWT}.
Introducing the coupling to the optical and mechanical baths responsible for energy dissipation, we obtain for the dynamics the following coupled %Quantum Langevin 
Dynamics Equations (DE) \cite{QLE}:
\begin{eqnarray}   
%\begin{gathered}  
\langle\dot{\hat{a}}\rangle &= &\left(+\mathbbm{i}\Delta'-\kappa_{tot}/2\right)\langle\hat{a}\rangle+\mathbbm{i}g_{0}\langle(\hat{b}+{\hat{b}}^{\dag})\hat{a}\rangle \label{alphabeta} \\ 
&-& \mathbbm{i} \Lambda_1 \langle(2\,{\hat{a}}^{\dag} \hat{a}+1 )\hat{a} \rangle \nonumber\\  
& - &\mathbbm{i} \Lambda_2 \langle(3\,{\hat{a}}^{\dag} \hat{a}\,{\hat{a}}^{\dag} \hat{a} + 3\,{\hat{a}}^{\dag} \hat{a}+1 )\hat{a} \rangle -\sqrt{\frac{\kappa_{ext}}{2}} \, \tilde{\alpha}_p e^{-\mathbbm{i}\varphi_p},\nonumber  \\
%\end{eqnarray}
%\begin{eqnarray}
\langle\dot{\hat{b}}\rangle &=&\left(-\mathbbm{i}\Omega_{m}-\Gamma_{m}/2\right)\langle\hat{b}\rangle+\mathbbm{i}g_{0}\langle{\hat{a}}^{\dag}\hat{a}\rangle , \label{betaalpha}
%\end{gathered}
\end{eqnarray}
see Appendix \ref{QLEcommute} for details.
In the limit of large field amplitudes, we can neglect quantum fluctuations and use the standard semiclassical approach:  $\langle\hat{a}\rangle\to\alpha$ and $\langle\hat{b}\rangle\to\beta$ (with all related replacements), leading to:
\begin{eqnarray}  
\dot{\alpha} &= &\left(+\mathbbm{i}\Delta'' -\kappa_{tot}/2\right)\alpha +\mathbbm{i}g_{0}(\beta+{\beta}^{*})\alpha  \label{eqsLangevin1} \\ 
&-& \mathbbm{i}\, 2 \tilde{\Lambda}_1 |\alpha|^2 \alpha -\mathbbm{i} \,3 \tilde{\Lambda}_2 |\alpha|^4\alpha -\sqrt{\frac{\kappa_{ext}}{2}} \, \tilde{\alpha}_p e^{-\mathbbm{i}\varphi_p}, \nonumber \\
\dot{\beta} &=&\left(-\mathbbm{i}\Omega_{m}-\Gamma_{m}/2\right)\beta +\mathbbm{i}g_{0} |\alpha|^2.  \label{eqsLangevin2}
\end{eqnarray}
The detuning and Kerr coefficients are recast into $\Delta''$, $\tilde{\Lambda}_1$ and $\tilde{\Lambda}_2$ respectively, which are defined in Appendix \ref{QLEcommute}.
We solve this system of coupled equations by means of the usual ansatz for $\beta$ \cite{CRphysiqueADA}:
\begin{equation} 
\beta=\beta_{c}+Be^{-\mathbbm{i}\phi} e^{-\mathbbm{i}\omega t}, \label{betac}
\end{equation}
$\beta_{c}$ being related to a (marginal) static deflection $x_c$ of the beam, and $B  e^{-\mathbbm{i}\phi}$ corresponding to the (complex valued) self-sustained motion. The oscillation frequency $\omega \sim \Omega_m$ will be defined self-consistently at the end of the Section.
Following Ref. \cite{marquardtBif}, we define 
$\alpha = \tilde{\alpha} e^{-\mathbbm{i}\Theta}$ and:
\begin{eqnarray}
%\begin{gathered}  
\Theta (t) & = & -z \sin(\omega t +\phi) ,\label{Jacobi}  \\
z & = &\frac{2 g_0 B}{\omega}. \nonumber
%\end{gathered}
\end{eqnarray}
This leads to the simpler set of equations:
\begin{eqnarray}  
\dot{\tilde{\alpha}} &= &\left(+\mathbbm{i}\Delta'' -\kappa_{tot}/2\right)\tilde{\alpha} -\sqrt{\frac{\kappa_{ext}}{2}} \, \tilde{\alpha}_p  e^{+\mathbbm{i}(\Theta-\varphi_p)} \label{eqsLangevin} \\ 
&-& \mathbbm{i}\, 2 \tilde{\Lambda}_1 |\tilde{\alpha}|^2 \tilde{\alpha} -\mathbbm{i} \,3 \tilde{\Lambda}_2 |\tilde{\alpha}|^4\tilde{\alpha}  ,   \nonumber \\
0 &=&\left[+\mathbbm{i}(\omega-\Omega_{m})-\Gamma_{m}/2\right] B \, e^{-\mathbbm{i}\phi} +\mathbbm{i}g_{0}  |\tilde{\alpha}|^2 e^{\mathbbm{i}\omega t} . \label{stabil}
\end{eqnarray}
In Eq. (\ref{stabil}), the term $|\tilde{\alpha}|^2 e^{\mathbbm{i}\omega t}$ is thus considered to be static; the lost non-resonant components are precisely the ones neglected by the chosen ansatz (which is an extremely good approximation for a high-Q mechanical mode, see Appendix \ref{AnsatzBetac}).
The related equation for $\beta_c$ is discussed in the same Appendix; rigorously speaking, one also has to replace $\Delta'' \rightarrow \Delta''+ 2 g_0 \Re [\beta_c]$. In practice, the static deflection is far too small to be of any relevance and can be safely neglected.

We look for a solution in the Fourier space. We therefore define $\tilde{\alpha}(t)=\sum_{n \in \mathbb{Z}} \tilde{\alpha}_n e^{+\mathbbm{i}n \omega t}$ (which leads to our comb), and re-write the problem at hand in terms of the complex amplitudes $\tilde{\alpha}_n$. Making use of the Jacobi-Anger transform, we obtain: 
\begin{eqnarray}
\tilde{\alpha}_n & = & \label{alphatilde} \\  
&& \!\!\!\!\!\!\!\!\!\!\!\!\!\!\!\!\!\!  \frac{+\sqrt{\frac{\kappa_{ext}}{2}   } \, \tilde{\alpha}_{p} J_n(-z) e^{+\mathbbm{i} (n \phi-\varphi_p)}+ A_n + B_n \vert \tilde{\alpha}_n\vert^2}{+\mathbbm{i}\left[\Delta'' - n \omega - D_n \right]- \mathbbm{i} C^{(1)}_n \vert \tilde{\alpha}_n\vert^2 - \mathbbm{i} C^{(2)}_n \vert \tilde{\alpha}_n\vert^4-\frac{\kappa_{tot}}{2}} , \nonumber \\ 
0 & = & \left[+\mathbbm{i}(\omega-\Omega_{m})-\Gamma_{m}/2\right] %\nonumber \\
%& + &  
+ \mathbbm{i}g_{0}  \frac{  \sum_{n \in \mathbb{Z}} \tilde{\alpha}_n\tilde{\alpha}_{n+1}^{*}}{B \, e^{-\mathbbm{i}\phi}} , \label{Beq}
\end{eqnarray}
with $Jn$ Bessel's function of first kind. 

The cavity field writes in the Fourier space $\alpha = \sum_{n \in \mathbb{Z}} \alpha_n e^{\mathbbm{i}n \omega t}$, with the $\alpha_n$
complex amplitudes deduced from the $\tilde{\alpha}_n$. 
Since the lock-in measurement has a narrow bandwidth around a chosen $n \omega$ frequency  \cite{DylanPRR}, the measured photon field amplitude $\dot{N}_{out,n}$ for each spectral peak $n$ is:
\begin{eqnarray} 
\dot{N}_{out,n} &=& \frac{\kappa_{ext}}{2} \vert \alpha_n\vert^2,  \\
\alpha_n & = &  e^{+\mathbbm{i}n \phi} \sum_{p \in \mathbb{Z}} J_{p-n}(-z) e^{-\mathbbm{i} p \phi}\tilde{\alpha}_p,  \label{Ndot}
\end{eqnarray}
where we made use, as for the input field, of the input-output relation \cite{devoret,DylanPRR}.
Again for a single-port system, remove the $1/2$.
With our notations, $n=+1$ corresponds to the Stokes sideband at $\sim \omega_c$ in the laboratory frame ($\sim -\Omega_m$ in the rotating frame, since $\omega \approx \Omega_m$ as demonstrated below) \cite{error}.

The coefficients $A_n, B_n, C^{(1)}_n, C^{(2)}_n$ and $D_n$ are given in Appendix \ref{AllCoeffs}.
They are defined from the amplitudes $\tilde{\alpha}_m$ with $m \neq n$, and involve a phase factor $\xi_n = \tilde{\alpha}_n^*/\tilde{\alpha}_n$. But they are independent of the amplitude $\vert \tilde{\alpha}_n\vert$: therefore, taking Eq. (\ref{alphatilde}) and multiplying it by its conjugate, one obtains a polynomial in $\vert \tilde{\alpha}_n\vert^2$ that can be solved, for each $n$, {\it knowing all the coefficients}.
For weak enough Kerr parameters, we can calculate  these coefficients from the linear expressions of $\tilde{\alpha}_n$ [Eq. (\ref{alphatilde}) with all nonlinear coefficients zero]. For larger Kerr terms, the approximation is not good enough and we can recalculate $A_n, B_n, C^{(1)}_n, C^{(2)}_n, D_n$ iteratively by reinjecting the obtained $\tilde{\alpha}_n$ into the expressions of Appendix \ref{AllCoeffs}. In practice, the procedure converges relatively quickly and typically 4 iterations of the process are enough. Besides, all sums $n \in \mathbb{Z}$ are truncated at $\pm N_{max}$; in practice $ N_{max}=4$ is a good compromise for precision versus calculation speed. 

When solving for $\vert \tilde{\alpha}_n\vert^2$, we obviously keep only the physical roots $>0$. But even then, the problem at hand can lead to {\it multiple optical solutions}, which is actually one of the features we are looking for (Appendix \ref{AllCoeffs}). When this occurs, we can only {\it extend by continuity} our original solution though this bifurcation point; the procedure cannot tell us whether the state found is stable or not, 
%%%% added referee
or {\it what type} of bifurcation is experienced when the system switches from one dynamic state to another.
%%%%
 The optical multi-stability zone is then represented in grey on the calculated plots.
Note that under a transformation $\tilde{\alpha}_n \rightarrow \tilde{\alpha}_n e^{+\mathbbm{i} (n \phi-\varphi_p)}$, the coefficients $C^{(1)}_n, C^{(2)}_n, D_n$ remain unchanged, while $A_n \rightarrow A_n e^{+\mathbbm{i} (n \phi-\varphi_p)}$ and $B_n \rightarrow B_n e^{+\mathbbm{i} (n \phi-\varphi_p)}$. This means that the phase factor next to $J_n(-z)$ in Eq. (\ref{alphatilde}) can actually be factorized out: as a result, it vanishes from Eqs. (\ref{Beq},\ref{Ndot}) which implies that both $\varphi_p, \phi$ are irrelevant (as they should). From now on, we shall drop them from the equations.

The $\tilde{\alpha}_n$ complex amplitudes depend on $B$ though $J_n(-z)$, Eq. (\ref{alphatilde}),
which in turn affects all nonlinear coefficients.
It now has to be found self-consistently, following the standard procedure \cite{CRphysiqueADA}. From Eq. (\ref{Beq}) we define:
\begin{eqnarray}
\Gamma_{BA}(B) & = & -2 \,\Re \left[\mathbbm{i}g_{0}  \frac{  \sum_{n \in \mathbb{Z}} \tilde{\alpha}_n(B)\,\tilde{\alpha}_{n+1}^{*}(B)}{B  } \right] ,\label{gba}  \\
\delta \Omega(B) & = & - \Im \left[\mathbbm{i}g_{0}  \frac{  \sum_{n \in \mathbb{Z}} \tilde{\alpha}_n(B)\,\tilde{\alpha}_{n+1}^{*}(B)}{B  } \right]. \label{deltaf}
\end{eqnarray}
In order to fulfill Eq. (\ref{Beq}), one then imposes $\Gamma_{BA}(B)/\Gamma_m+1=0$ which fixes the $B$ amplitudes, and then $\omega=\Omega_m+\delta \Omega(B)$ the actual mechanical resonance frequencies ($\vert \delta \Omega \vert \ll \Omega_m$, and can be neglected when computing the fields $\tilde{\alpha}_n$). When more than one value of $B$ verifies the condition, one obtains {\it mechanical multistability} \cite{DylanPRR}. This is commented in Appendix \ref{MechHysteresisSolve}, and such regions appear in yellow in the theoretical plots.

\begin{figure}[t!]
		\centering
	\includegraphics[width=9cm]{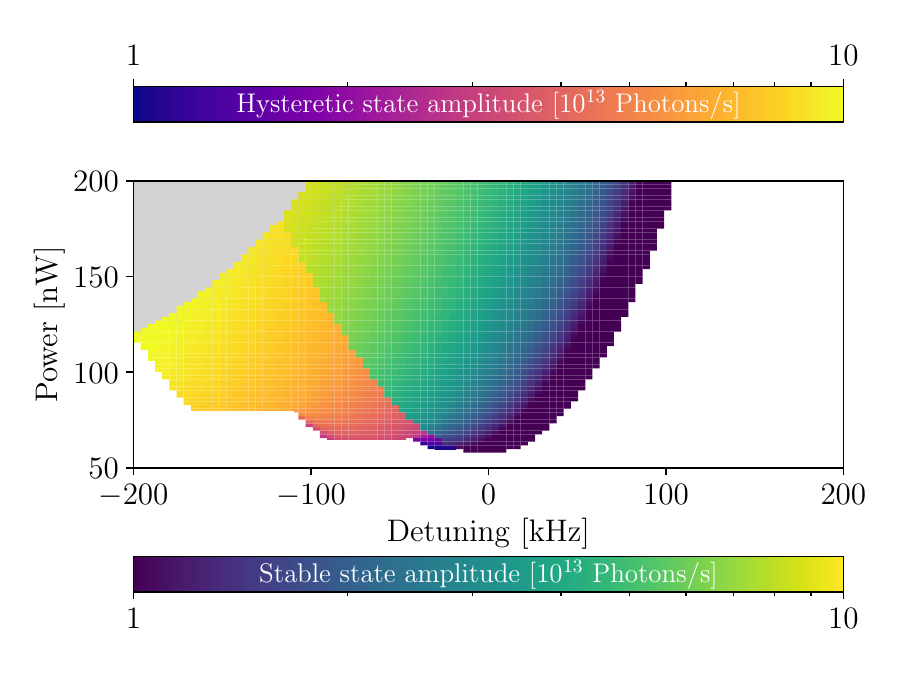}
	%\vspace*{-0.5cm}
			\caption{
			%(Color online) 
			Calculated plot corresponding to Fig. \ref{fig2} from the presented Kerr theory. The color coding and the scale bars are identical, reproducing the measured features (see text for details).}
			\label{fig3bis}
\end{figure}

\section{Role of Kerr nonlinearities}	
\label{results}

The calculated amplitude parameter space for the peak at frequency $\sim \omega_c$ is presented in 
 Fig. \ref{fig3bis}.
 This is our best fit, adjusting {\it only} the two Kerr coefficients to $\tilde{\Lambda}_1 \approx - 5.2 \times 10^{-4}~$Rad/s and $\tilde{\Lambda}_2 \approx +9.9 \times 10^{-16}~$Rad/s. The cavity frequency  corrections $\delta \omega', \delta \omega''$  (Appendix \ref{HamiltonRWT}, \ref{QLEcommute}) are negligible, and set to zero. The agreement is fairly good: it reproduces rather accurately Fig. \ref{fig2}, with the grey color corresponding to the optical multistability region (Appendix \ref{AllCoeffs}), and the yellow one to the mechanical hysteresis (the experimental sweeping procedure has been taken into account to produce the "staircase-like" bottom, see Appendix \ref{DuffingShiftg1g2}). Quantitatively, the predicted amplitude is of the right magnitude, and the grey region at the right place; however, the boundaries of the yellow region in the $\delta < 0$ half-space are only approximate. %, but {\it does not} present any moir\'e.
%%%%
%The green region appears to be perfectly symmetric around $\delta \sim 0$, which is {\it not} the case in Fig. \ref{fig2} (resulting in a much smaller yellow region on the theory plot).
%%%%
As well, the other comb amplitudes seem to fall faster with detuning than the measured ones (see Fig. \ref{fig1}; note that the peak around $\omega_c+\Omega_m$ also contains the pump signal).
%%%%%
The theory is therefore not perfect, even though we have been trying to address all sources of deviations we could possibly think of (e.g. static deflection in Appendix \ref{AnsatzBetac}, Duffing nonlinearity in Appendix \ref{DuffingShiftg1g2}, or thermal gradients in Appendix \ref{HeatingandT}, all of which being irrelevant as far as Fig. \ref{fig3bis} is concerned).
%%%%%
%%%% MODIF!!!
Only the nonlinear coupling terms have not been included so far, and they {\it must} be relevant to some extent, as demonstrated in Ref. \cite{DylanPRR}.
They are explicitly addressed in Appendix
\ref{DuffingShiftg1g2}, and discussed in the following. \\
%%%%%

The presented theory has two fit parameters $\tilde{\Lambda}_1, \tilde{\Lambda}_2$, the leading orders of our Kerr expansion.
It is enlightening to consider the impact of each of these terms on the calculation.
Setting $\tilde{\Lambda}_2=0$, and adjusting {\it only} $\tilde{\Lambda}_1$, one can create a mechanical hysteresis (yellow zone in Fig. \ref{fig3bis}), and reduce both photon and motion amplitudes. 
However, no optical multistability (grey zone) can be produced. Such a region appears {\it only} when setting a finite value to $\tilde{\Lambda}_2$; but with a small enough parameter, photon amplitudes and hysteresis are not much affected. It thus appears that the effect of the two coefficients is rather different: essentially, adding higher order Kerr terms such as $\tilde{\Lambda}_2$ mostly increases the size of the solution space [the order of the polynomial to solve, Eq. (\ref{alphatilde})], but does not modify noticeably the solution which we track by continuity.
%%%
This justifies our truncation of the Kerr expansion at order 2.
%%%
Similarly, the nonlinear coupling coefficients  cannot create an optical instability such as the grey zone, and do not impact much the overall photon amplitude; but they can modify the mechanical hysteresis (see Appendix \ref{DuffingShiftg1g2}). This actually means that $\tilde{\Lambda}_1$ can be fit almost independently on the overall height of the photon signal: and it has to be negative to decrease the amplitude as compared to the linear theory. As a result, the mechanical hysteretic region (yellow) emerges at the proper position in the parameter space (on the left).
%%%% Comment fit!!!
Presumably, adjusting the nonlinear coupling coefficients should allow matching precisely the shape of this yellow region; %, and create the characteristic moir\'e
 but this implies to fine tune a large enough set of nonlinear coupling coefficients, a very tedious task which does not influence our main conclusions
 (see Appendix \ref{DuffingShiftg1g2}). This is clearly outside of the scope of the present manuscript. \\
%%%%

The coefficient $\tilde{\Lambda}_1$ can be evaluated from circuit parameters \cite{alessandro}. It writes \cite{DylanPRR,mis2}:
\begin{equation}
\tilde{\Lambda}_1 = - \frac{\alpha_l \, \hbar \omega_c^2}{L_g \, (2/3)^3 I_c^2 } ,
\end{equation}
with $\alpha_l = L_k/L_g$ the ratio of kinetic inductance to the geometric one ($L_g$ about $1.3~$nH for our circuit), and $I_c = J_c \, A_{ef\!f}$ the critical current (flowing at $\sim 6~$GHz, in an effective cross section $A_{ef\!f}$). The kinetic inductance (for cavity length $l \sim 3~$mm) is defined by \cite{tinkham}:
\begin{eqnarray}
L_k &=& \frac{m^*}{2 \, n_s A_{ef\!f} \,e^2} \, l , \\
n_s &=& \frac{J_c}{e \,v_c} ,
\end{eqnarray}
with $e$ the electric charge, $m^*$ the effective charge carrier mass, and $v_c=\Delta_0/p_F$ the critical velocity. The latter can be evaluated from $\Delta_0 =1.76 \lambda_{BCS} \, k_B T_c$ (well known BCS-gap equation) and $p_F=\sqrt{2 m^* E_F}$ ($E_F$ the Fermi energy, tabulated) \cite{tinkham,ashcroft}.
From the literature, we  chose $\lambda_{BCS} \approx 1.5$ (strong coupling correction) and $m^*=2\, m_0$
($m_0$ the electron mass) \cite{gap,mass}.
  The critical temperature $T_c$ of our niobium layer has been measured to be about 9$~$K on a test sample, as well as the critical current density $J_c \sim 5 \times 10^{10}~$A/m$^2$ (using a DC current). Both values are compatible with published results on Nb thin films \cite{supra}. Using twice the width of the central conductor of the coplanar cavity meander to evaluate $A_{ef\!f}$ (the AC current profile flows in the inner conductor, but also partly in the ground), one obtains precisely the fit $\tilde{\Lambda}_1$.
As a rule of thumb, one could guess 
$\vert \tilde{\Lambda}_2 \vert \sim \omega_c \left( \tilde{\Lambda}_1/\omega_c\right)^2 \sim  10^{-17}~$Rad/s, which is actually much smaller than the fit value.
And precisely, the grey-yellow/green border in Fig. \ref{fig3bis}
is extremely sensitive to $\tilde{\Lambda}_2$: one can argue that it offers a unique way of experimentally
%%%% MODIF
 assessing the weight of higher order Kerr coefficients, here limited to order 2 in our expansion.
%%%
Finally, coming back to the work of Cattiaux {\it et al.} \cite{DylanPRR} we verify that one needs a Kerr coefficient $\Lambda_1 \sim 10^{-2}~$Rad/s in order to impact their theoretical modeling, far larger than the expected one for their device. This validates the fact that the Kerr nonlinerity could be neglected in Ref. \cite{DylanPRR}, focusing on the nonlinear coupling terms.
%%%%

A final discussion concerning the amplitude of motion is in order. The gap $d$ between the beam device and its electrode, Fig. \ref{fig1} right inset, is designed to be about 100$~$nm. Measuring carefully with an SEM, we find out that it is 140$~$nm in the center, and increases to 160$~$nm at the clamps. The calculation predicts an amplitude of motion as large as 80$~$\% of this gap. This is remarkably close to a collapse of the beam into the nearby wall, which would actually happen {\it in absence of Kerr nonlinearities}: it turns out that by decreasing the amplitude of motion, $\tilde{\Lambda}_1$ actually protects the device from destruction.

\begin{figure*}[t!]	 
\center
			 \includegraphics[width=19.cm]{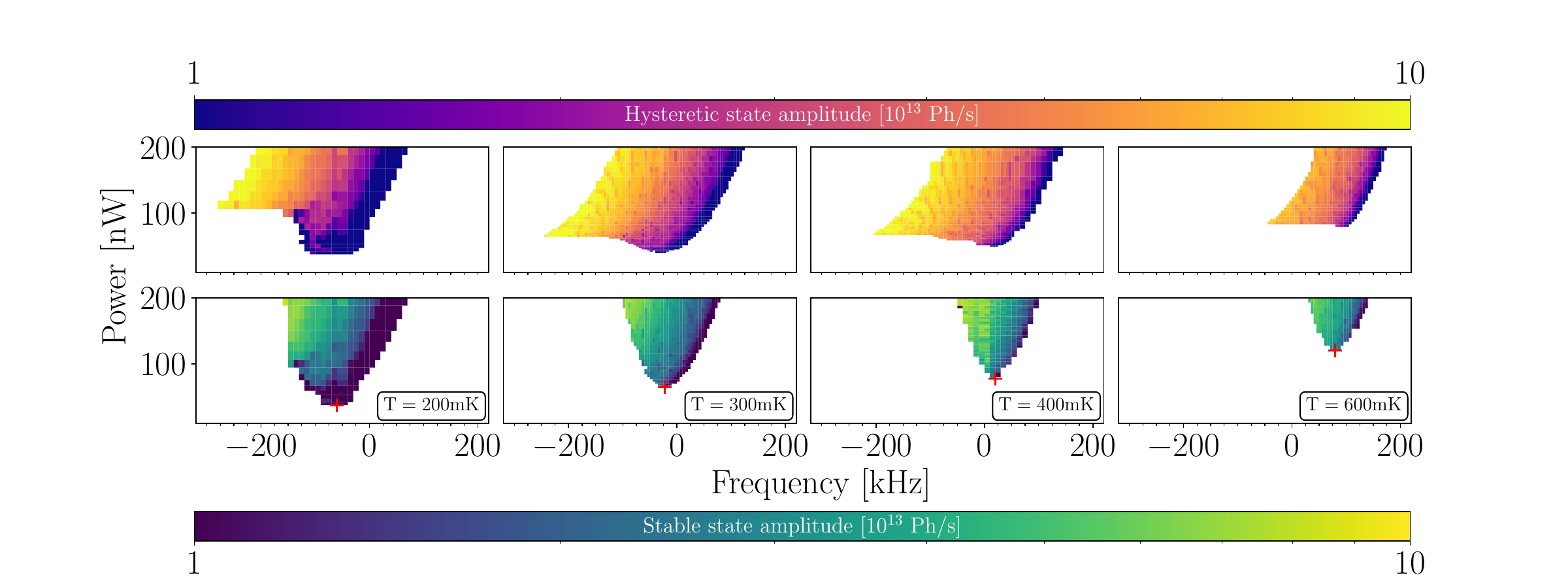}
			\caption{
			Measured amplitude of the main comb peak in the parameter space $(P_{in},\delta)$ at different temperatures. 
%%%% XXX MODIF
   From left to right: 200$~$mK to 600$~$mK (see legend). Color coding identical to other plots (the grey zone on the top lefts is omitted for clarity):
%%%%% XXX MODIF			
 here, the yellow maps are measured with horizontal (detuning) scans while the green ones are obtained from vertical (power) scans, demonstrating hysteresis. Note the slight irreproducibility in the patterns, obtained here by stitching multiple measurements. The red crosses mark the entry in the self-oscillation regime at the actual zero detuning % , while the grey bar on the left graph represents the typical threshold reproducibility in power 
 (see text for discussion).}
			\label{figA}
		\end{figure*}
		
\section{Conclusion}

%%%%% ON a plus que la conclu à faire!!!!!!

%Duffing standard $\beta_D == 3 \lambda_D/(4 x_{zpf}^2)$ Expect Duffing $10^{18}~$Hz/m$^2$. So lambdaD about 1e-9

%%% elsewhere !! grey-yellow/green  EQUATIONS BREAk Appendix

%XXXXX to do!!
%
%Can we measure zizi? Appendix \ref{MechHysteresisSolve}.
%Comment Reproducibility on diff cooldowns?
%Measure other comb map?
%
%Redo the SIMU for the drum, with lambda1=1e-4!!
%Validates no effect? Hum.
%
%Plus the gn in appendix?
%creates wiggles????
%
%impact of Duffing and g1 g2 ... gn??? Can we speculate on their magnitude, estimate??
%%%%%%% 
%And g1 g2 g3 in Appendix; extra plot with them and boundary values? Or just discuss???? Also in Results Section???
%%%
%Discuss Appendix \ref{DuffingShiftg1g2}... And here???
% FWHH 5.5 Hz peak fig 1; beyond current understanding...

We report on the limit cycle dynamics of a nanomechanical beam device coupled to a high-Q superconducting coplanar cavity. By applying a strong blue-detuned microwave tone, the mechanical mode enters into self-oscillation (parametric instability).
%%%%%%%%
We find out experimentally that in a given range of the power-detuning parameter space, this coherent optomechanical dynamic state is inevitably annihilated: only a "power on/off" action can revive the coupled system.
%%%%%%%%%
We develop the theory based on the cavity Kerr nonlinearity that describes this striking result.
The agreement between the model and the measurements is semi-quantitative, catching the main features and predicting the proper photon amplitudes. 
%%%%
The fit Kerr coefficients are in reasonable agreement with expectations. 
%%%% ADDED!
They appear to have opposite signs, as if there was a sort of "saturation" of the nonlinear phenomenon. The physical reason behind this fact remains unknown.
%%%%
The main imprint on the dynamics is the appearance in the parameter space of an {\it optical instability}: the photon population becomes multi-valued.
%%%%%
We therefore interpret our remarkable experimental finding as a consequence of a bifurcation in the intra-cavity microwave field towards a very high population state.
This state should in turn be responsible for very large currents in the conducting layer that destroy superconductivity. This transition is highly hysteretic, and requires the drive field power to be decreased significantly in order to restore the initial state. \\
%%%%%%

A complete quantitative modeling of the system should also include nonlinear coupling coefficients \cite{DylanPRR}. %; this shall be addressed elsewhere.
%%%%
 These terms are certainly responsible for %the moir\'e pattern observed experimentally, and the slight asymmetry of the stable (green) region 
the slight disagreement between theory and experiment in the exact location of the % two stability 
hysteretic zone
 (compare yellow range in Fig. \ref{fig2} and Fig. \ref{fig3bis}).
% g1 g2 g3 future work quantitatif et moire?
%%%%% ADD energy transfer
The study of the {\it full comb} (and not only the main peak) might also lead to new insights. In the present work, we could not identify any energy leakage from one comb peak to another, or any secondary bifurcations visible only in secondary peaks; but this would deserve to be clarified and carefully studied in the future.
 
%%%%%
Besides, modeling all the optical branches, and defining their respective stability range remains a task to be done. This is outside of the capabilities of our analytic approach, and would certainly require numerical resources.
%%%%%%%%%%%%%%%%%%%%%%%%
In particular, a time domain analysis of the model (together with dedicated measurements) could help unraveling the nature of the bifurcations. For instance, it might be possible to further clarify the nature of the hysteretic region and the coexisting attractors involved. A similar analysis has been performed e.g. in Ref. \cite{canard}.
%%%%%%%%%%%%%%%%%%%%%%%%

We believe that our work gives a good understanding of the measured self-sustained motion above 200$~$mK; it can also be used as an efficient tool to evaluate the  Kerr nonlinear coefficients, 
%%% Modif!
being quantitative on the first one $\tilde{\Lambda}_1$. 
%%%%
The complete knowledge of the system's Hamiltonian (including the nonlinear terms) is indeed of the utmost importance for some fundamental research areas \cite{quantgravitnonlin,tobias_SB}.
%%%%
However, the coupled optomechanical dynamics {\it below} this temperature becomes much more complex, and does not fit anymore in the present theoretical framework. The threshold from Brownian motion towards self-oscillation at zero detuning becomes strongly hysteretic with respect to injected microwave power \cite{XinPRAppl}; multiple bifurcations can be identified, whose origin is still unknown. 
%%%%
While tentative explanations are addressed in Appendices, these fundamental issues are calling for further developments. \\
%%%%%%%%

On the applied side, this specific feature that destroys a microwave mode could be particularly relevant as a new resource for quantum electronics. 
For instance, it could be seen as the implementation of a "microwave fuse", which switches off a signal irreversibly through a tiny modification of a control parameter (but reversibly, without destroying the chip). Alternatively, one could optimize the setup to create an "edge detector", which would very sensitively switch to a {\it no signal state} when the optical bifurcation occurs.
%\newpage

%%%% XXXXXX ADDED DISCUSSION!

\vspace*{1cm}
(\dag) Corresponding Author: eddy.collin@neel.cnrs.fr

\begin{acknowledgements}

The Authors acknowledge the use of the N\'eel {\it Cryogenics} and {\it Nanofab} facilities. 
E.C. would like to thank  A. Armour, M. Dykman, I. Favero and H. Cercellier for very useful discussions. As well, A.D. thanks A. Metelmann for enlightening discussions.
We acknowledge support from the ERC StG grant UNIGLASS No. 714692 (A.F.), and the French National Research Agency (ANR), grant ANR-MORETOME, No. ANR-22-CE24-0020-01 and No. ANR-22-CE24-0020-02 (X.Z. and E.C.).
The research leading to these results has received funding from the European Union's Horizon 2020 Research and Innovation Programme, under grant agreement No. 824109, the European Microkelvin Platform (EMP).

\end{acknowledgements}

\appendix
		
\section{T-dependence and heating}
\label{HeatingandT} 
		
Experiments have been conducted in the range $100~$mK $-$ 1$~$K. Measured maps of the photon flux at frequency $\sim \omega_c$ as a function of $(P_{in},\delta)$ are shown in Fig. \ref{figA}, for a set of temperatures.

All of these plots look fairly similar, apart from a $T$-dependent frequency and power shift. This can be 
quantitatively illustrated with the crosses in Fig. \ref{figA}, located at the entrance to the self-oscillation regime with lowest $P_{in}$.
As the temperature rises, $\Gamma_m$ increases linearly (as $0.1\times T$[mK] in Hz), and so does $P_{in}$ at this point. Concomitantly, the cavity frequency $\omega_c$ shifts upwards (in this range essentially as $+300\times T$[mK] in Hz, while the mechanical frequency moves by no more than about $+20~$Hz which can be neglected). This results in an apparent increase in the fit $\Delta=\Omega_m+\delta$ minimum position because of our chosen fixed reference based on the value of $\omega_c$ at $T=380~$mK (where the minimum of the map is truly at $\delta=0$). These temperature dependencies are fairly consistent with the ones of earlier works on the same device \cite{XinPRAppl}.
%%%%% CORR
The graphs have been acquired by pieces (nominally, the two half-spaces $\delta < 0$ and $\delta >0$). Stitching them together creates tiny irregularities: it shows that the patterns have some irreproducibility in them, typically about $\pm 0.5~$dB and $\pm 10~$kHz at 400$~$mK (and seemingly getting worse as we cool down). 
%%%% MODIF
Besides, a small hysteresis exists between green and yellow zones {\it even for $\delta \geq 0$}, which is not captured by our model. This could be due to a material-related nonlinear mechanism that would lead to slightly larger mechanical damping rates in the low-motion state (Brownian) compared to the large-motion one (self-oscillation).
 Alternatively, coupling nonlinear coefficients could be the origin of this, see following Appendix \ref{DuffingShiftg1g2}.
 %%%% ADDD!!!
Interestingly, after re-setting the system when the optical instability has "destroyed" the dynamic state (namely recovering from the grey zone of the plots), the mechanical damping rate observed is systematically slightly {\it smaller} than initially (e.g. about $15~\%$ at 400$~$mK). It takes hours for the device to relax towards the original $\Gamma_m$, which contributes to the irregularities in the measured patterns. One could interpret this fact as a signature of weakly coupled internal degrees of freedom within the beam (potentially, the TLSs) being highly out of equilibrium (and saturated), and slowly recovering their thermal state. This would be in agreement with a rather high temperature reached by the sample upon switching the superconducting metal to its normal state.
%%%%

Certainly the most remarkable feature of these temperature measurements is that the coldest set of data is at 200$~$mK: the self-oscillation regime looks drastically different at lower temperatures. More complex patterns are seen, presumably with {\it more} stable states and bifurcation points. 
%%%% Add
This could be due to a particularly wiggly stability equation $\Gamma_{BA}/\Gamma_m+1=0$ (commented in Appendix \ref{DuffingShiftg1g2} in the framework of nonlinear coupling).
%%%
The irreproducibility seems also to increase.
%%%%
This shall be the subject of further studies, and we can simply for the time being point out that 200$~$mK is precisely the temperature where "spikes" anomalies have been reported in the Brownian regime \cite{XinPRAppl}. 

We shall now comment on the signatures of heating, due to microwave absorption in the dielectric substrates. This effect was rather strong for the device studied in Ref. \cite{DylanPRR}; while here it is very mild.
Drifts in the temperature of the mechanical element have been characterized in the Brownian regime in Ref. 
\cite{XinPRAppl}. The reported number in our temperature range is about $\Delta T$[mK] $ \sim 0.1 \, P_{in}$[nW]. Extrapolated to the self-oscillation range at higher powers, this would lead at most to a temperature increase of about 20$~$mK, leading to at most 10$~\%$ change in $\Gamma_m$ and a marginal mechanical frequency shift.
On the other hand, the cavity power dependence is dominated by internal mechanisms linked to the physics of Two Level Systems (TLSs) \cite{XinPRAppl}: as we increase $P_{in}$, TLSs get saturated and the damping $\kappa_{tot}$ improves (while the frequency $\omega_c$ drifts down), until we reach a limit where both are stable, at powers slightly below the self-oscillation threshold. On top of this, no heating effect can be established: the power dependence would indeed be {\it the opposite} (damping increase, frequency shift up).
This can be expected: the thermal anchoring of the cavity to the chip has to be good, compared to the one of the mechanical device, since it is quite large and not suspended. We have to conclude that this is enough to counteract the microwave absorption, which is obviously larger in the cavity than in the beam alone.
Extrapolating this result to the self-oscillation regime, we expect cavity heating to be a marginal effect in our experiments. At worst, $\omega_c$ would drift linearly upwards with $P_{in}$, at a very small pace. 

But such tiny temperature drifts in $\Gamma_m$ and $\omega_c$ could easily be taken into account in 
Fig. \ref{figA} as fit parameters: their only effect would be to slightly distord our calculated maps by pushing them at high powers toward the top-right. By no means could it explain the overall shape of our parameter space, and certainly not the multi-stabilities.
This has to be contrasted to experiments performed with laser optomechanics, where thermo-optical contributions are observed (potentially leading to bistabilities) \cite{painterBif,thermikFavero}.

\section{Mechanical nonlinearities}
\label{DuffingShiftg1g2} 
		
Mechanical nonlinearities have been studied in Ref. \cite{DylanPRR}. They are ubiquitous, and therefore also exist in the device we studied experimentally here.
While the measured features are clearly dominated by the Kerr (optical) terms,  
in the present Appendix we shall consider their impact on our calculations.

We consider first the Duffing effect, which is due to the stretching of the beam under motion \cite{duffingbook}: formally, we add to the Hamiltonian a term $\hat{H}_D = +\hbar \lambda_D \left({\hat{b}}+{\hat{b}}^{\dag} \right)^4/4$. 
The Duffing coefficient $\lambda_D$, expressed here in Rad/s, shall be discussed below.
The dynamics equation Eq. (\ref{eqsLangevin2}) of the mechanics is then altered with the adjunction of a term  $-\mathbbm{i} \lambda_D \left( \beta+\beta^{*} \right)^3 $ on the {\it rhs}; the optical field dynamics expression is unmodified.
		
We solve by adopting the ansatz Eq. (\ref{betac}). As such, we develop the new term in Eq. (\ref{eqsLangevin2}) keeping only components that are static or oscillating as $\exp[-\mathbbm{i} \omega t]$. Other terms impact only the negligible non-resonant components of the mechanical motion, as discussed in Appendix \ref{AnsatzBetac}. We obtain:
\begin{eqnarray}
-\mathbbm{i} \lambda_D \left( \beta+\beta^{*} \right)^3 & = & -\mathbbm{i} \lambda_D \left[12 B^2+ 8\Re(\beta_c)^2 \right] \Re(\beta_c) \label{newDuffing}  \\
& & -3 \mathbbm{i} \lambda_D \left[ B^2 +4 \Re(\beta_c)^2 \right] B e^{-\mathbbm{i}\phi} e^{-\mathbbm{i}\omega t}  . \nonumber
\end{eqnarray}
The first term in Eq. (\ref{newDuffing}) modifies the static component solution Eq. (\ref{staticB}):
\begin{eqnarray}
\Re \left[ \beta_c \right] & = & \frac{g_0 \Omega_m \sum_{n \in \mathbb{Z}} \vert \tilde{\alpha}_n\vert^2 - 8 \lambda_D \Omega_m \Re \left[ \beta_c \right]^3}{\Omega_m^2 +12 \Omega_m \lambda_D B^2+\Gamma_m^2/4}. \label{DuffBetac}
\end{eqnarray}
The cubic term $\Re \left[ \beta_c \right]^3$ is actually very small and can be safely neglected. However, we find out that the harmonic amplitude $B$ potentially impacts the static displacement $x_c= x_{zpf} 2 \Re \left[ \beta_c \right]$ (introducing the zero point fluctuation $x_{zpf}$), see denominator of Eq. (\ref{DuffBetac}): for a "hardening" Duffing effect $\lambda_D >0$ (which is the case for doubly-clamped beams), the static displacement is {\it reduced} by a large motion amplitude $B$. However numerically, this is found to be marginal. 

\begin{figure}[t!]
		\centering
	\includegraphics[width=8.5cm]{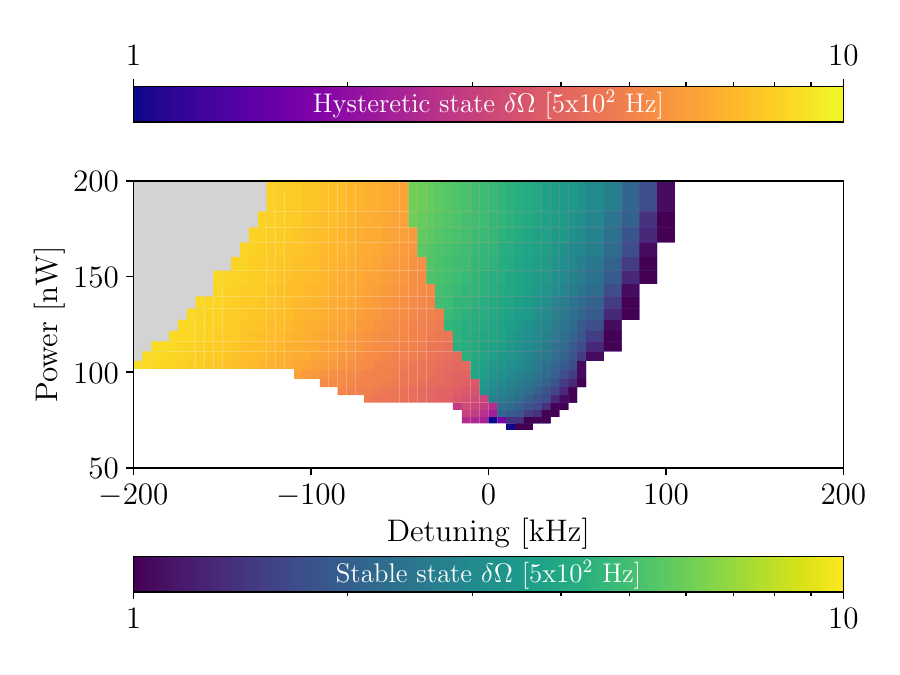}
	%\vspace*{-0.5cm}
			\caption{
			%(Color online) 
			Measured mechanical frequency shift (obtained from the detected peak position) at 400$~$mK. The color coding is the same as for the other plots (see text). }
			\label{figB1}
\end{figure}

\begin{figure}[t!]
		\centering
	\includegraphics[width=8.5cm]{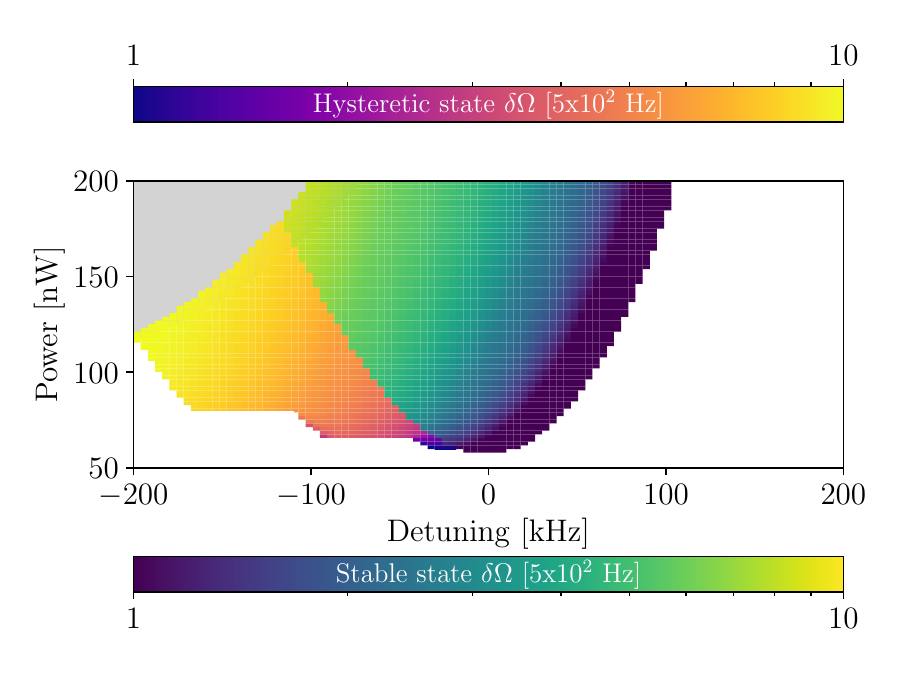}
	%\vspace*{-0.5cm}
			\caption{
			%(Color online) 
			Calculated mechanical frequency shift corresponding to Fig. \ref{figB1} (identical color coding, see text for details). }
			\label{figB2}
\end{figure}

The second term in Eq. (\ref{newDuffing}) modifies the stability equation Eq. (\ref{Beq}). But it 
only adds a pure imaginary term, which does not alter our definition of $\Gamma_{BA}$. It can be conveniently absorbed in Eq. (\ref{deltaf}) by posing:
\begin{equation}
\delta \Omega(B) \rightarrow \delta \Omega(B) + 3 \lambda_D \left[ B^2 +4 \Re(\beta_c)^2 \right].
\end{equation}
It is again perfectly safe to neglect $\Re \left[ \beta_c \right]$ in the above, which means that the Duffing effect simply adds a {\it frequency shift} $\propto B^2$ to the mechanical resonance \cite{DylanPRR}. 

\begin{figure*}[t!]	 
\center
			 \includegraphics[width=18.cm]{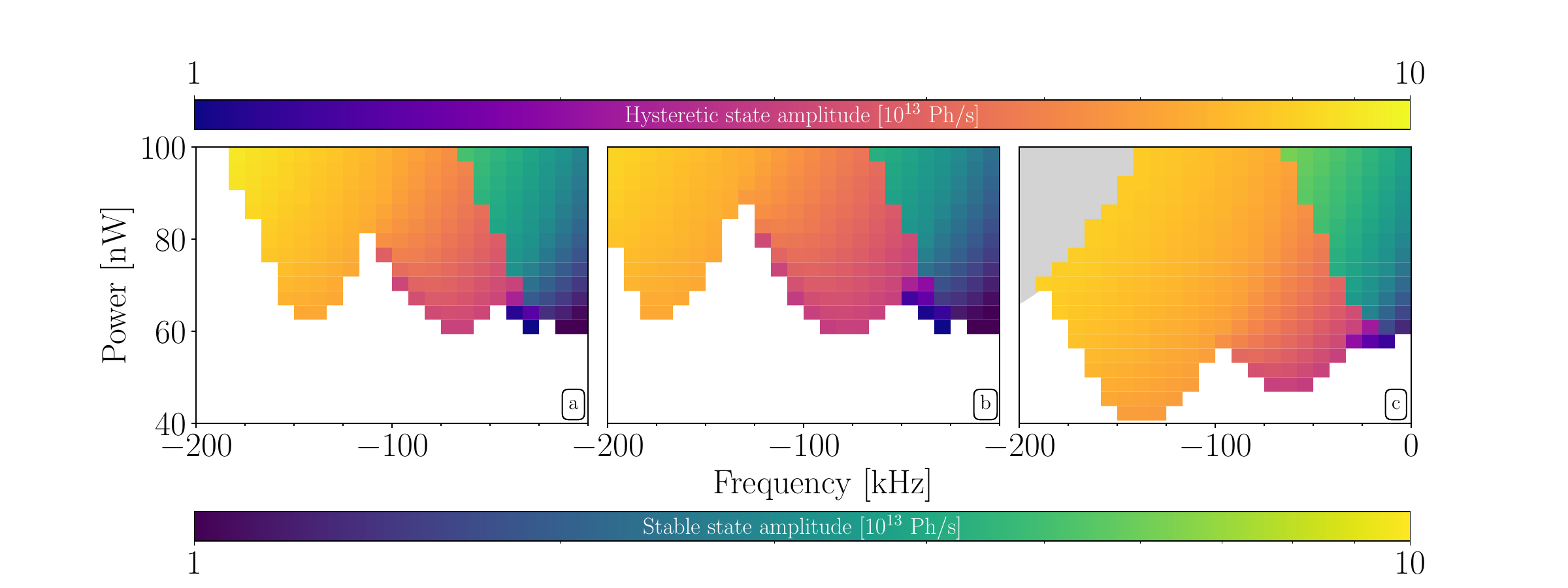}
			\caption{
			a): Zoom-in on the hysteretic region of the theoretical plot Fig. \ref{fig3bis}, without taking into account the sweep procedure ($g_1= g_2=0$, see text). b): Calculation performed by adding $g_1= 5. 10^{-8}~$Hz , $g_2 =0$ to the model: the "spiky" structure shifts to the left. c): Calculation performed with $g_1= 0 $, $g_2 =+3. 10^{-14}~$Hz: this time, the "spiky" structure shifts down. Reversing signs of $g_1$ and $g_2$, these structures seem to displace the other way. The rest of the plot (especially the amplitudes) is essentially unaltered (see text). }
			\label{figG3}
		\end{figure*}

The measured mechanical frequency shift corresponding to the data in Fig. \ref{fig2} is shown in Fig. \ref{figB1}. The calculated shift corresponding to Fig. \ref{fig3bis}, using $\lambda_D \approx +2 \times 10^{-9}~$Rad/s as a fit parameter, is displayed in Fig. \ref{figB2}. The colors are based on the same convention as in the core of the paper (green stable, yellow mechanical bistability, and grey optical multistability).

The agreement is fairly good, and converting $\lambda_D$ into conventional units for the Duffing pulling effect, we find $+2.8 \times 10^{17}~$Hz/m$^2$, which matches reasonably expectations for our beam device \cite{duffingbook,martial}. 
Besides, the maximal frequency shifts are about a few kHz, which has to be compared to $\Omega_m/(2\pi) \sim 4~$MHz: this justifies the assumption $\vert \delta \Omega\vert  \ll \Omega_m$ in the solving of the stability equation.
Note also that in these conditions, the Duffing effect {\it can by  no means} lead to mechanical bistability (as is however the case for a standard driven resonator): any hysteresis in the mechanical response (as discussed in Appendix \ref{MechHysteresisSolve}) has to be explained by other mechanisms. \\

%%%%%%% HEEEERE
The second mechanical nonlinear feature is in the coupling strength \cite{DylanPRR}: while in the standard theory the capacitive coupling is limited to the first order term $\partial C/\partial x$ (Appendix \ref{HamiltonRWT} below), with very large motion amplitudes higher orders (e.g. $\partial^2 C/\partial x^2$, $\partial^3 C/\partial x^3$) come into play.
This can be taken into account with non-linear coupling coefficients $g_1, g_2$ which modify the expressions Eqs. (\ref{alphatilde},\ref{Ndot}) by replacing the Bessel functions with $J_n(-z) \rightarrow f_n(z)$. The function $f_n(z)$ is explicitly given in Ref. \cite{DylanPRR}, and an extension to higher (arbitrary) orders can be found in Ref. \cite{DylanPhD}.

%%%%%% Discuss g1 g2
Including $g_1$ and $g_2$ nonlinear coefficients in the theory, we find out that the calculations are not altered for $g_1 \in [ - 10^{-8} ; +10^{-8} ]$ and $g_2 \in [-10^{-14} ; +10^{-14}]$ (in Hz). 
%%%
Note that for each extra $g_n$ term added to the algorithm, the required calculation power grows substantially.
%%%
With $g_n$ $(n=1,2)$ extending slightly out of these limits, the calculated stability function $\Gamma_{BA}/\Gamma_m+1$ which defines the dynamic state (Appendix \ref{MechHysteresisSolve}) gets progressively distorted. As a result, their initial impact is to modify the borders of the hysteretic region (yellow zone), {\it without modifying substantially} neither the optical multistability range (grey zone) nor the overall amplitudes. 
An example of this is given in Fig. \ref{figG3}, comparing the theoretical graph of Fig. \ref{fig3bis} with the one obtained using finite values $g_1\neq 0 , g_2 \neq 0$. 
Here, we did not take into account the sweeping technique which creates the very flat bottom of the $\delta < 0$ region (as in Ref. \cite{DylanPRR}); instead, we see that the theory produces characteristic "spiky" shapes at the bottom of the yellow zone. 
%This obviously means that a simple $\delta$ scan as the one produced in the experiment cannot trigger these zones, which are cut out and simply absent from Fig. \ref{fig2}: this creates a very straight border at the bottom of the yellow region in Fig. \ref{fig3bis}, as in the theoretical plots of Ref. \cite{DylanPRR}. 
These structures are very sensitive to the $g_n$ values, as they can shift both horizontally and vertically (Fig.  \ref{figG3}). But we found that measuring them carefully with a specific sweeping profile in the $(P_{in},\delta)$ space proves to be essentially impossible: the system is not stable enough at that resolution level (see Appendix \ref{HeatingandT}). 

Besides, specific arrangements of $g_n$ parameter values can create theoretically not only a "dip" in the stability function (see Fig. \ref{figG2}), which is responsible for mechanical hysteresis, but can also create a "peak" just below it.
Such a peak means that, in addition to our large motion amplitude hysteretic state, {\it there must exist another self-oscillating state at much lower motion amplitudes}. This would explain the asymmetry of the measured green region in Fig. \ref{fig2}, since entering it from the left $\delta < 0$ would require to overcome this peak height (requiring thus a higher $P_{in}$). On the $\delta >0$ side of the parameter space, none of these features (dip or peak) appears, meaning that the border of the green zone matches the linear calculation (Fig. \ref{fig3}).
Trying to explore this effect, we could not experimentally observe such a low amplitude self-oscillation state above 200$~$mK, which might mean that it remains too small. However, this mechanism based on nonlinear coupling could explain the appearance of new dynamic states at lower temperatures, as briefly commented in Appendix \ref{HeatingandT} and Conclusion: the asymmetry of the green zone would then be an indirect signature of these. 
%%%%
%Note that the Kerr effect clearly masks nonlinear coupling signatures: a calculation performed with $\tilde{\Lambda}_{1}=0, \tilde{\Lambda}_{2}=0$ shows that a sole $g_1= xxx$ does create a mechanical hysteresis on the left part of the parameter space (but no optical one), while with Kerr terms it has essentially no impact.
On the other hand, when $g_n$ terms are much larger than the limits specified above, the calculated self-oscillation map is completely modified and does not resemble anymore Fig. \ref{fig2}. It obviously means that they have to be experimentally small enough to remain within the reasonable limits explored in the present Appendix.

One can actually guess what these reasonable limits must be in practice. From dimensional analysis, we expect $g_n/g_0 \sim (x_{zpf}/d)^n$, with $d$ the gap between the beam and the electrode (and $x_{zpf}$ the zero point fluctuation). Numerically, $x_{zpf}/d \sim 2. \times 10^{-7}$, which does (within a factor ten) correspond to $g_n$ values {\it that would influence the calculated pattern, without being too large}.
%%%%%
It is therefore perfectly reasonable to argue that neglecting nonlinear coupling does capture the new measured features, especially the optical instability. As well, the discussion of the present Appendix justifies why the modeling cannot be perfectly quantitative without the $g_n$ terms, and why the complexity of the full nonlinear problem imposes to leave them outside of the present work.

\section{Gereric Hamiltonian and Rotating-Wave Transform}
\label{HamiltonRWT} 

We consider in the first place the following Hamiltonian (in the laboratory frame):

\begin{eqnarray}
\hat{H} & = & \hbar \omega_c \left( {\hat{a}}^{\dag} {\hat{a}} + \frac{1}{2} \right)+ \hbar\Omega_{m} \left( {\hat{b}}^{\dag}\hat{b} + \frac{1}{2} \right) \label{hamil} \\
& & -\hbar g_{0}\left(\hat{b}+{\hat{b}}^{\dag}\right) {\hat{a}}^{\dag}\hat{a}  \nonumber \\ 
& & -\mathbbm{i} \hbar \sqrt{\frac{\kappa_{ext}}{2}} \left[ {\hat{a}}^{\dag}  \alpha_{p} (t)  -\hat{a}\, {\alpha_p}^{*}(t)  \right] \nonumber \\
&  & + \frac{\hbar \lambda_3}{3} \left( \hat{a}+ {\hat{a}}^{\dag} \right)^3 + \frac{\hbar \lambda_4}{6} \left( \hat{a}+ {\hat{a}}^{\dag} \right)^4 \nonumber \\
& & + \frac{\hbar \lambda_5}{10} \left( \hat{a}+ {\hat{a}}^{\dag} \right)^5 + \frac{\hbar \lambda_6}{20} \left( \hat{a}+ {\hat{a}}^{\dag} \right)^6 + \cdots \nonumber .
\end{eqnarray}
The first line corresponds to the uncoupled optical (${\hat{a}}^{\dag}, {\hat{a}}$ creation/anihilation operators) and mechanical (${\hat{b}}^{\dag}, {\hat{b}}$) fields, %%%changed!!
and the second line to their coupling (with $g_0 =x_{zpf} \, \partial \omega_c/\partial x $).  $x_{zpf}=\sqrt{\hbar/(2 m_0 \Omega_m)}$ is the zero point fluctuation of the mechanical mode (with $m_0$ its mass), and $\partial \omega_c/\partial x$ quantifies the coupling strength arising from the motion $x$ of one cavity mirror \cite{AKMreview}. For microwave optomechanics, this end mirror is actually a capacitor \cite{regal2008}, and $\partial \omega_c/\partial x = \partial \omega_c/\partial C \times \partial C/\partial x$ with $\partial \omega_c/\partial C=-\omega_C/(2 C_0)$ and $C_0$ the cavity mode capacitance (for us $C_0 \approx 0.5~$pF). The third line corresponds to the coupling to the optical drive (through $\kappa_{ext}/2$ for a bidirectional coupling), $\alpha_p(t)$ being the sinusoidal microwave tone \cite{devoret}. 

The two last lines stand for the cavity nonlinearity. Following Ref. \cite{metelmann}, 
we assume here a nonlinear expansion in terms of the electric field amplitude $\propto ( {\hat{a}} + {\hat{a}}^{\dag})$, and we push it to sixth order. The $\lambda_i$ are thus the generic Kerr coefficients that describe our nonlinear superconducting cavity; the fit values are discussed in Section \ref{results} of the paper. 

Since the time scales of the two resonators are so well separated, and the Q factors are high, the first step of the calculation consists in applying a Rotating Wave Transform (at the frequency of the pump $\omega_p$), neglecting all fast rotating terms. In this rotating frame, Eq. (\ref{hamil}) writes (dropping also the irrelevant constant terms): 
\begin{eqnarray}  
%\begin{gathered}  
\hat{H} & = &-\hbar  \Delta \,{\hat{a}}^{\dag}\hat{a} -\hbar g_{0}\left(\hat{b}+{\hat{b}}^{\dag}\right) 
 {\hat{a}}^{\dag}\hat{a} + \hbar\Omega_{m} {\hat{b}}^{\dag}\hat{b} \label{HofRWT} \\ 
&  &\!\!\!\!\!\!\!\!\!\!\!\!\!\!\!\!\!  -\mathbbm{i} \hbar \sqrt{\frac{\kappa_{ext}}{2}} \left[ {\hat{a}}^{\dag} \tilde{\alpha}_{p} e^{-\mathbbm{i}\varphi_p} -\hat{a}\, \tilde{\alpha_p}^{*}e^{+\mathbbm{i}\varphi_p} \right]  \nonumber \\
&  &\!\!\!\!\!\!\!\!\!\!\!\!\!\!\!\!\! + \hbar \lambda_4 \left( {\hat{a}}^{\dag}\hat{a} \right) \left( {\hat{a}}^{\dag}\hat{a} \right)  + \hbar \lambda_4 \left( {\hat{a}}^{\dag}\hat{a} \right) \nonumber \\
& &\!\!\!\!\!\!\!\!\!\!\!\!\!\!\!\!\! +\hbar \lambda_6 \left( {\hat{a}}^{\dag}\hat{a} \right) \left( {\hat{a}}^{\dag}\hat{a} \right) \left( {\hat{a}}^{\dag}\hat{a} \right) +\hbar \frac{3 \lambda_6}{2} \left( {\hat{a}}^{\dag}\hat{a} \right) \left( {\hat{a}}^{\dag}\hat{a} \right)+\hbar \frac{\lambda_6}{2} \left( {\hat{a}}^{\dag}\hat{a} \right) . \nonumber
%\end{gathered}
\end{eqnarray}
The odd orders have disappeared, and we introduced the complex drive amplitude $\tilde{\alpha}_{p} e^{-\mathbbm{i}\varphi_p}$ and $\Delta = \omega_p-\omega_c$. This expression leads to Eq. (\ref{hamilton}) when we define:
\begin{eqnarray}
\Delta' & = & \omega_p - \omega_c' , \\
\omega_c' & = & \omega_c + \delta \omega ' , \\
\delta \omega ' & = & \lambda_4 +\frac{\lambda_6}{2}  , \\
\Lambda_1 & = & \lambda_4 +\frac{3 \lambda_6}{2} , \\
\Lambda_2 & = & \lambda_6.
\end{eqnarray}
Note that one could have postulated directly a Taylor expansion of the Hamiltonian in terms of photon population ${\hat{a}}^{\dag}\hat{a}$ (instead of field amplitude). Then, the nonlinear terms involving $\Lambda_1, \Lambda_2$ in  Eq. (\ref{hamilton}) would be identical to the original ones defined in the laboratory frame, since they remain unaltered by the Rotating Wave Transform. As such, it also simplifies $\delta \omega' = 0$ in the above, or equivalently $\Delta' = \Delta$.

\section{Dynamics Equations} %Quantum Langevin Equations
\label{QLEcommute} 

From Eq. (\ref{hamilton}), one has to generate the dynamics equations for ${\hat{a}}$ and ${\hat{b}}$ operators.
In the Heisenberg picture, one writes $\dot{{\hat{a}}} = -\mathbbm{i}/\hbar \left[{\hat{a}}, \hat{H} \right]$ and $\dot{{\hat{b}}} = -\mathbbm{i}/\hbar \left[{\hat{b}}, \hat{H} \right]$, making use of  the Hamiltonian Eq. (\ref{HofRWT}) to derive Eq. (\ref{hamilton}).
Besides, to be complete we must include the surrounding optical and mechanical baths, 
which shall lead to the definition of the damping terms $-\kappa_{tot}/2 \, {\hat{a}}$ and $-\Gamma_{m}/2 \, {\hat{b}}$ respectively \cite{QLE}. We leave this aspect out of our discussion, and the interested reader can consult the aforementioned reference on open quantum systems.

What nonlinearities bring to the calculation is thus terms of the type $\propto \left[{\hat{a}}, {\hat{N}}^n \right]$, with ${\hat{N}} = {\hat{a}}^{\dag}{\hat{a}}$ the photon number operator and $n>1$ an integer. We have:
\begin{eqnarray}
\left[{\hat{a}}, {\hat{N}}^2 \right] & = & \left( 2 \hat{N} +1 \right) {\hat{a}}, \\
\left[{\hat{a}}, {\hat{N}}^3 \right] & = & \left( 3 \hat{N}^2+3 \hat{N} +1 \right) {\hat{a}},
\end{eqnarray}
which then lead to the result
 Eqs. (\ref{alphabeta},\ref{betaalpha}) presented in the core of the paper.
Terms can be regrouped in order to obtain a more compact writing, as used in Eq. (\ref{eqsLangevin1}). We thus define:
\begin{eqnarray}
\Delta '' & = & \Delta ' - \Lambda_1 - \Lambda_2=\omega_p-\omega_c'', \\
\omega_c'' & = & \omega_c +\delta \omega'' , \\
\delta \omega '' & = & \delta \omega'+\Lambda_1 + \Lambda_2 = 2\lambda_4 +3 \lambda_6  , \\
\tilde{\Lambda}_1 & = &  \Lambda_1 + \frac{3  \Lambda_2}{2} = \lambda_4+ 3 \lambda_6,\\
\tilde{\Lambda}_2 & = & \Lambda_2 = \lambda_6,
\end{eqnarray}
which are the relevant parameters to be fit; see Section \ref{results}.

\section{Static deflection and out-of-resonance terms}
\label{AnsatzBetac} 

The ansatz Eq. (\ref{betac}) is an extremely good approximation of the exact mechanical solution (a frequency comb that derives from the one imprinted in the optical field):
\begin{eqnarray}
\beta & = &\beta_{c}+B e^{-\mathbbm{i}\phi} e^{-\mathbbm{i}\omega t}  \\
&& + \sum_{m<0}  B_m e^{-\mathbbm{i}(m \phi+\epsilon_m)} e^{-\mathbbm{i} m \omega t} \nonumber \\
&& + \sum_{p>1}  B_p e^{-\mathbbm{i} (p\phi+\epsilon_p)} e^{-\mathbbm{i}p \omega t} \nonumber .
\end{eqnarray}
The terms $B_m, B_p$ (and their related phases $\epsilon_m, \epsilon_p$) correspond to the non-resonant amplitudes of the mechanical motion.
We anticipate that their impact shall be very small, and the aim of Appendix \ref{AnsatzBetac} is to demonstrate it mathematically.
This expression leads to a coupling term in Eq. (\ref{eqsLangevin1}) of the form:
\begin{eqnarray}
&&+\mathbbm{i} g_0 (\beta+\beta^*) \alpha  =  +\mathbbm{i} \, 2 g_0 \Re \left[ \beta_c \right] \alpha \label{truesol} \\
& & + \mathbbm{i} \, 2 g_0 \bar{B} \cos(\omega t + \bar{\phi}) \, \alpha  \nonumber \\
&& + \mathbbm{i} \sum_{p>1} 2 g_0 \bar{B}_p \cos(p \omega t + \bar{\phi}_p) \,  \alpha, \nonumber
\end{eqnarray}
where we have defined:
\begin{eqnarray}
\bar{B} & \approx & B + B_{-1} \cos \left(\epsilon_{-1} \right),\\
\bar{\phi} & = & \phi + \bar{\epsilon} , \\
\bar{\epsilon} & \approx & - \frac{B_{-1}}{B} \sin \left(\epsilon_{-1} \right)  ,
\end{eqnarray}
for the main term oscillating at frequency $\omega$, and similarly for $p>1$:
\begin{eqnarray}
& & \!\!\!\!\! \bar{B}_p  =  \sqrt{B_{p}^2+B_{-p}^2 +2 \, B_{p} B_{-p} \cos \left(\epsilon_{-p}+\epsilon_p \right) },\\
& &\!\!\!\!\! \bar{\phi}_p  = p \phi + \bar{\epsilon}_p , \\
& &\!\!\!\!\!   \bar{\epsilon}_p   = \arctan \left[ \frac{- B_{-p} \sin \left( \epsilon_{-p} \right) + B_{p} \sin \left( \epsilon_{p}\right) }{B_{-p} \cos \left( \epsilon_{-p} \right) + B_{p} \cos \left( \epsilon_{p}\right)} \right] \label{epsilonphase} \\
& & \,\,\,\,\, + mod \, \pi . \nonumber
\end{eqnarray}
In the above, we already took into account that the amplitude $B_{-1}$ is much smaller than $B$.
The $mod$ term in Eq. (\ref{epsilonphase}) is the conventional $0, \pm 1$ depending on the signs of numerator and denominator in the bracket expression (see trigonometry rules). 

The first term in Eq. (\ref{truesol}) is nothing but a cavity frequency shift, which can be absorbed into $\omega_c'' \rightarrow \omega_c'' -2 g_0 \Re \left[ \beta_c \right] $. The second term is our usual amplitude modulation, while the last one corresponds to all the non-resonant components. Each of these has formally the same structure as the main one, and one can still simplify the dynamics equation by means of multiple applications of the Jacobi-Anger transform. One defines $\alpha = \tilde{\alpha} e^{-\mathbbm{i}\Theta_{tot}}$:
\begin{eqnarray}
\Theta_{tot}(t) & = & \Theta(t) +\sum_{p>1} \Theta_p(t) , \\
\Theta_p(t) & = & - z_p \sin \left( p \omega t+p \phi + \bar{\epsilon}_p \right) , \\
 z_p & = & \frac{2 g_0 \bar{B}_p}{p \omega},
\end{eqnarray}
similarly to Eq. (\ref{Jacobi}). In the writing of $\Theta(t)$, we make the replacement $B \rightarrow \bar{B}$ and $\phi \rightarrow \bar{\phi}= \phi + \bar{\epsilon}$. Eq. (\ref{eqsLangevin}) is then simply modified by $\Theta \rightarrow \Theta_{tot}$, and Eq. (\ref{alphatilde}) has to be modified with $J_n(-z) \rightarrow f_n(z)$.
Truncating the $\Theta_{tot}$ sum at $p=3$, the function $f_n$ writes:
\begin{eqnarray}
f_n(z) & =  & \sum_{p,m \in \mathbb{Z}} \left(-e^{-\mathbbm{i}\bar{\epsilon}_3} \right)^m \left( -e^{-\mathbbm{i}\bar{\epsilon}_2}\right)^p
 e^{+\mathbbm{i} (3m+2p+n) \bar{\epsilon}} \nonumber\\
& & \times J_m (-z_3) J_p(-z_2) J_{3m+2p+n}(-z),\label{newexpansion}  
\end{eqnarray}
which can be extended straightforwardly to higher orders.

Each of the complex amplitudes $B e^{-\mathbbm{i}{\phi}}$, $B_k e^{-\mathbbm{i}{\phi}_k}$ (with $k>1$ or $ k<0$) and $\beta_c$  is defined from Eq. (\ref{stabil}). The main term still writes as Eq. (\ref{Beq}), by construction, while for the others we have:
\begin{eqnarray}
\Re \left[ \beta_c \right] & = & \frac{g_0 \Omega_m \sum_{n \in \mathbb{Z}} \vert \tilde{\alpha}_n\vert^2 }{\Omega_m^2 +\Gamma_m^2/4},  \label{staticB} \\
B_k e^{-\mathbbm{i}{\epsilon}_k} & = & \frac{\mathbbm{i} g_0 \, e^{+\mathbbm{i}k{\phi}}\sum_{n \in \mathbb{Z}}   \tilde{\alpha}_n \tilde{\alpha}_{n+k}^* }{+\mathbbm{i} (\Omega_m - k \omega) +\Gamma_m/2} . \label{allBk}
\end{eqnarray}
Note that for $\beta_c$, only the real part has a physical meaning which is why Eq. (\ref{staticB}) is expressed this way. For $B_k$ however, the expressions are complex-valued; but the main phase $\phi$ shall cancel out in all the sums written above, see Section \ref{theory}. 

When computing the stability equation $\Gamma_{BA}/\Gamma_m+1$ from Eq. (\ref{gba}), using the function $f_n(z)$ instead of the usual $J_n(-z)$, 
the problem becomes in principle far more complex: now one has to define $B$ {\it and all} the $B_k$ self-consistently. 
However in practice, the $B_k$ are so small that one can proceed iteratively. We first set all $B_k$ to zero, recovering from 
Eq. (\ref{newexpansion}) the original problem that depends {\it only} on $B$. We can then use Eq. (\ref{allBk}) to calculate the amplitudes and phases $B_k, \epsilon_k$, and re-inject them in Eq. (\ref{newexpansion}). It turns out that the found solution is not altered to a very high accuracy, which demonstrates its validity.

We shall illustrate this with values corresponding to our experiment. Eq. (\ref{staticB}) corresponds to a static mechanical displacement of the beam $x_c=x_{zpf}\, 2 \Re \left[ \beta_c \right]$. 
For our device, this is of order $1~$\AA ngstr\"om (to be compared to $x_{zpf}\, 2B$ of the order of 100$~$nm, Section \ref{results}), 
leading to a cavity shift of order 1$~$kHz (to be compared to $\kappa_{tot}$ of order hundreds of kHz, Section \ref{expt}).
The amplitudes $z_p$ scale inversely with the mechanical Q factor, and here they are at most $10^{-6}$ times smaller than the main contribution $z \propto B$. This clearly demonstrates that they can all be safely neglected, and that the ansatz Eq. (\ref{betac}) captures all what is relevant to describe the physics. This conclusion holds for any experimentally relevant set of parameters (namely for reasonably good mechanical Q factors).

\section{Self-Consistent nonlinear coefficients}
\label{AllCoeffs} 

The coefficients appearing in the definition of $\tilde{\alpha}_n$ in Section \ref{theory} involve sums on $\tilde{\alpha}_m$ for $m \neq n$, and a phase factor $\xi_n = \tilde{\alpha}_n^*/\tilde{\alpha}_n$.
They decompose into:
\begin{eqnarray}
A_n & = & + 2 \mathbbm{i} \tilde{\Lambda}_1 \, a_{1,n} + 3 \mathbbm{i} \tilde{\Lambda}_2 \, a_{2,n}, \label{An} \\
B_n & = & + 3 \mathbbm{i} \tilde{\Lambda}_2 \, b_{2,n}, \\
C_n^{(1)} & = & + 2  \tilde{\Lambda}_1 + 3  \tilde{\Lambda}_2 \, c_{2,n} , \\
C_n^{(2)} & = & + 3  \tilde{\Lambda}_2 , \\
D_n & = & + 2  \tilde{\Lambda}_1 \, d_{1,n} + 3  \tilde{\Lambda}_2 \, d_{2,n} , \label{Dn}
\end{eqnarray}
with:
\begin{eqnarray}
a_{1,n} & = & \!\!\!\!\!\!\!\!\!\!\!\! \sum_{\substack{\mbox{$k,p \neq n$} \\ \mbox{and $k+p \neq 2n$}}} \!\!\!\! \tilde{\alpha}_k \tilde{\alpha}_p \tilde{\alpha}_{k+p-n}^*, \\
d_{1,n} & = & 2 \sum_{p \neq n} \vert \tilde{\alpha}_p \vert^2 + \xi_n \sum_{p \neq n} \tilde{\alpha}_{p}  \tilde{\alpha}_{2n-p} , \\
a_{2,n} & = &  \!\!\!\!\!\!\!\!\!\!\!\!\!\!\!\!\!\!\!\!\! \sum_{\substack{\mbox{$k,p,q,m \neq n$} \\ \mbox{and $k+p+q-m \neq 2n$}}} \!\!\!\!\!\!\!\!\!\!\!\!\!\!\!\! \tilde{\alpha}_k \tilde{\alpha}_p \tilde{\alpha}_q \tilde{\alpha}_{m}^* \tilde{\alpha}_{k+p+q-m-n}^*, 
\end{eqnarray}
\begin{eqnarray}
b_{2,n} & = & 3 \!\!\!\!\!\!\!\!\sum_{\substack{\mbox{$k,p \neq n$} \\ \mbox{and $k+p \neq 2n$}}} \!\!\!\!\!\!\!\!\tilde{\alpha}_k \tilde{\alpha}_p \tilde{\alpha}_{k+p-n}^* \\
& &\!\!\!\!\!\!\!\! + 3 \xi_n^* \!\!\!\!\!\!\!\!\sum_{\substack{\mbox{$p,m \neq n$} \\ \mbox{and $p-m \neq 0$}}} \!\!\!\!\!\!\!\!\tilde{\alpha}_p \tilde{\alpha}_m^* \tilde{\alpha}_{p-m+n}^*  + 3   \sum_{ k,p,m \neq n } \tilde{\alpha}_k \tilde{\alpha}_p \tilde{\alpha}_{m}^* \nonumber \\
& & \!\!\!\!\!\!\!\!+  \xi_n \sum_{ k,p,q \neq n } \tilde{\alpha}_k \tilde{\alpha}_p \tilde{\alpha}_{q} , \nonumber \\
c_{2,n} & = & 6 \sum_{p \neq n} \vert \tilde{\alpha}_p \vert^2 \\
&& + 3 \xi_n \sum_{p \neq n}   \tilde{\alpha}_p \tilde{\alpha}_{2n-p} +\xi_n^* \sum_{p \neq n} \tilde{\alpha}_p^* \tilde{\alpha}_{2n-p}^*, \nonumber \\
d_{2,n} & = & 3 \!\!\!\!\!\!\!\!\!\!\!  \sum_{\substack{\mbox{$k,p,m \neq n$} \\ \mbox{and $k+p-m \neq n$}}} \!\!\!\!\!\!\!\!\!\!\! \tilde{\alpha}_k \tilde{\alpha}_p \tilde{\alpha}_{m}^* \tilde{\alpha}_{k+p-m}^* \\
& + & 2 \xi_n  \!\!\!\!\!\!\!\!\! \!\!\!\!\!\!\!\!\!  \sum_{\substack{\mbox{$k,p,q \neq n$} \\ \mbox{and $k+p+q \neq 3n$}}} \tilde{\alpha}_k \tilde{\alpha}_p \tilde{\alpha}_{q} \tilde{\alpha}_{k+p+q-2n}^*. \nonumber
\end{eqnarray}
Starting from the vector $\left\{ \tilde{\alpha}_n \right\}$, with $-N_{max} \leq n \leq + N_{max}$  calculated in the linear case, we can evaluate all the above expressions. Injecting these in Eq. (\ref{alphatilde}), we obtain a new vector $\left\{ \tilde{\alpha}_n \right\}$ which is used to recalculate them. The procedure is reiterated up to ten times, evaluating the distance between the 
$  \tilde{\alpha}_n  $ terms from one step to the other. We find out that this distance becomes extremely small within a few iterations, typically 4. It is thus enough to obtain a good evaluation of the whole vector $\left\{ \tilde{\alpha}_n \right\}$, taking into account the Kerr nonlinearities. \\

Two subtleties have to be pointed out. First, the whole procedure is realized {\it for a given} mechanical amplitude B. We therefore calculate the vector $\left\{ \tilde{\alpha}_n \right\}$ in a range of $B$, which shall be enough to solve the stability equation Eq. (\ref{Beq}), see Appendix \ref{MechHysteresisSolve} below.
%%%%
Second, when the nonlinear Kerr terms are large enough, Eq. (\ref{alphatilde}) leads to a set of solutions: the problem is multi-valued.
The number of physical roots $>0$ depends on the order of the polynomial in $\vert \tilde{\alpha}_n \vert^2$ created, namely on the number of Kerr terms we chose to keep in the expansion (and their magnitudes). 
Since our truncation is arbitrary (and relies only on a convergence argument), we cannot infer {\it how many} solution branches actually exist. We can only state that there is more than one.
%%%%
 When that happens, in the algorithm we chose the root which is the closest to the previous calculation: this produces a solution branch (within the grey zone in Fig. \ref{fig3bis}) which is {\it continuous} in the $(P_{in},\delta)$ parameter space. It essentially means that our Eq. (\ref{alphatilde}) is actually valid {\it only} for this solution. We are not able, with this procedure, to calculate the proper equation leading to the other optical branches. Besides, we cannot infer either if the solution found is stable or not. And precisely, our claim is that in the multi-valued range, the optical field switches to another branch (with presumably much higher amplitude).

The (complex-valued) coefficients Eqs. (\ref{An}-\ref{Dn}) have each a specific impact on the $\tilde{\alpha}_n$ expression. The real part of $D_n$, appearing at the denominator of Eq. (\ref{alphatilde}), is simply a frequency shift, while its imaginary part will modify the damping term $\kappa_{tot}$. At the numerator, $A_n$ is a renormalization of the amplitude (in both magnitude and phase), while $B_n$ appears in factor of $\vert \tilde{\alpha}_n \vert^2$: this corresponds to a non-linear drive amplitude. Finally, the two Kerr terms appear at the denominator with $C_n^{(1)}$ for the quadratic one, and $C_n^{(2)}$ for the quartic. Note that conceptually, the whole procedure described in this paper could be pushed to higher orders; however, the calculation of the related coefficients becomes outrageously difficult.
The relevance of each of these terms is discussed in Section \ref{results}.

\begin{figure}[t!]
		\centering
	\includegraphics[width=8cm]{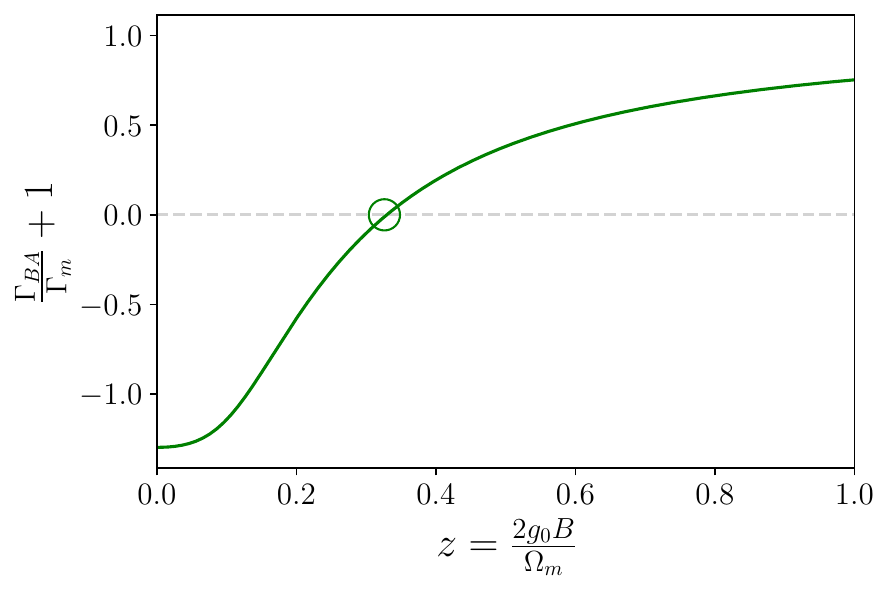}
	%\vspace*{-0.5cm}
			\caption{
			%(Color online) 
			Stability plot calculated at 400$~$mK for $P_{in}=130~$nW and $\delta/(2\pi) =0~$Hz, with the fit parameters discussed in Section \ref{results}. Only one stable state is found (circled point, see text). }
			\label{figG1}
\end{figure}

\section{Stability equation solving}
\label{MechHysteresisSolve} 

\begin{figure}[t!]
		\centering
	\includegraphics[width=8cm]{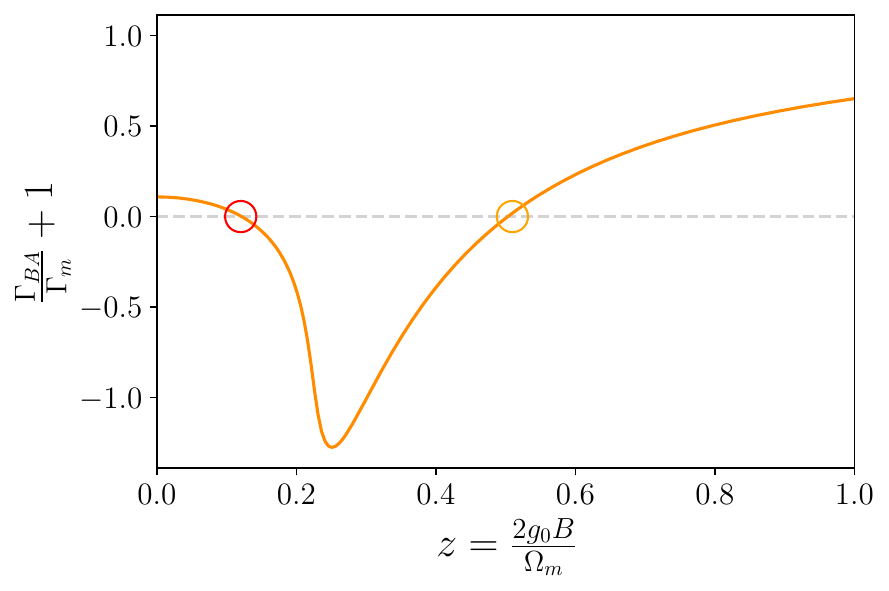}
	%\vspace*{-0.5cm}
			\caption{
			%(Color online) 
			Stability plot calculated at 400$~$mK for $P_{in}=130~$nW and $\delta/(2\pi) =-100~$kHz, with the fit parameters discussed in Section \ref{results}. A minimum in the curve leads to {\it mechanical hysteresis}, see text. }
			\label{figG2}
\end{figure}

The stability equation for the mechanical amplitude $B$, Eq. (\ref{Beq}), is solved numerically following the same procedure as in Ref. \cite{DylanPRR}. The only difference lies in the fact that here, we need a full set $\left\{ \tilde{\alpha}_n \right\}$ in order to compute it.
The small frequency shifts are neglected (including the Duffing effect, see Appendix \ref{DuffingShiftg1g2}), imposing $\omega = \Omega_m$ in the calculation to a very good accuracy.
Examples are shown in Figs. \ref{figG1} and \ref{figG2}. 

The situation shown in Fig. \ref{figG1} is the simplest one, and the most common: the function $\Gamma_{BA}/\Gamma_m+1$ obtained from Eq. (\ref{gba}) cuts the $x-$axis in only one place, which is graphically our solution for $B$. For the situation displayed, we obtain $\delta \Omega = +2\pi \times 134~$Hz from Eq. (\ref{deltaf}), which indeed verifies $\vert \delta \Omega \vert /\Omega_m \ll 1$ (as for all other points in the parameter space). 

But a more involved situation can occur: the function $\Gamma_{BA}/\Gamma_m+1$ can present a minimum, as shown in Fig. \ref{figG2}.
In this case, two intersections with the $x-$axis exist, and only the second one is stable (positive derivative, orange circle). 
This defines a $B$ state of very large amplitude that cannot be triggered from the thermal motion $B \sim 0$: it can be reached only from an already established large $B$ state, tuning $P_{in}$ or $\delta$ towards this state \cite{DylanPRR}. This is what creates a {\it mechanical hysteresis}, illustrated in yellow in the amplitude/frequency plots.
It turns out that the Kerr nonlinearity introduced 
in the modeling does precisely generate such mechanical hysteresis, by strongly impacting the function $\Gamma_{BA}$, as can be seen in Fig. \ref{figG2}. But the nonlinear coupling terms do as well (and they can even create more wiggly patterns at low $z$, see discussion in Appendix \ref{DuffingShiftg1g2}), which means that the actual fitting of the zone borders $\delta < 0$ produced in Fig. \ref{fig3bis} is particularly involved (and outside of our scope).
%%%% addd!
 The influence of the measurement sweeping technique on the definition of the stability zone borders is also addressed in the same Appendix.
% see Section \ref{results} for discussion. 

The calculation procedure is repeated for all $(P_{in},\delta)$ points of the map. The retained set $\left\{ \tilde{\alpha}_n \right\}$ which corresponds to the stable value of $B$ is then used to calculate the measured photon flux, Eq. (\ref{Ndot}).
Mechanical frequency shifts are discussed in Appendix \ref{DuffingShiftg1g2}.

\end{document}